\documentclass[pre,aps,reprint,longbibliography,superscriptaddress]{revtex4-2}

\usepackage{tikz}
\usepackage{amsmath}
\usepackage{graphicx}
\usepackage{float}
\usepackage{color}
\usepackage{comment}
\usepackage{bbold,bm}
\usepackage{relsize}
\usepackage[caption=false]{subfig}
\usepackage{gensymb}
\usepackage{siunitx}
\usepackage{float}
\usepackage{scalerel,amssymb}

\usepackage{pifont}

\newcommand{\kB}[0]{k_{\rm B}}
\newcommand{\rvec}[0]{{\bm r}}

\newcommand{\K}[0]{{\sqrt{\epsilon/m\sigma ^2}}}
\newcommand{\tunit}[0]{{\sqrt{m\sigma ^2/\epsilon}}}
\usepackage{soul}


\usepackage[english]{babel}
\usepackage[utf8]{inputenc}
\usepackage{psfrag} 
\usepackage{epstopdf}

\begin{document}

\title{Binary Mixtures of Locally Coupled Mobile Oscillators}

\author{Gon\c{c}alo Paulo}
\affiliation{Departamento de F\'{\i}sica, Faculdade de Ci\^{e}ncias, Universidade de Lisboa, 
    1749-016 Lisboa, Portugal}
    \affiliation{Centro de F\'{i}sica Te\'{o}rica e Computacional, Faculdade de Ci\^{e}ncias, Universidade de Lisboa, 1749-016 Lisboa, Portugal}

    \author{Mykola Tasinkevych}
    \email{mtasinkevych@fc.ul.pt}
    \affiliation{Departamento de F\'{\i}sica, Faculdade de Ci\^{e}ncias, Universidade de Lisboa, 
    1749-016 Lisboa, Portugal}
    \affiliation{Centro de F\'{i}sica Te\'{o}rica e Computacional, Faculdade de Ci\^{e}ncias, Universidade de Lisboa, 1749-016 Lisboa, Portugal}

\begin{abstract}

We study synchronization dynamics in binary mixtures of locally coupled Kuramoto oscillators which perform Brownian motion in a two-dimensional box. We introduce two models, where in model $\cal I$ there are two type of oscillators, say $\cal A$ and $\cal B$, and any two similar oscillators tend to synchronize their phases, while any two dissimilar ones tend to be out of phase. In model $\cal J$, in contrast, the oscillators in subpopulation $\cal A$  behave as in model $\cal I$, while the oscillators in subpopulation $\cal B$ tend to be out of phase with all the others. In the real space all the oscillators in both models interact via a soft-core repulsive potential. Both subpopulations of model $\cal I$ and subpopulation $\cal A$  of model $\cal J$, by their own, exhibit a phase coherent attractor in a certain region of model parameters. The approach to the attractor, after an initial transient regime, is exponential with some characteristic synchronization time scale $\tau$. Numerical analysis reveals that the attractors of the two subpopulations survive within model $\cal I$, regardless of the composition of the mixture $\phi$ and the strength of the cross-population negative coupling constant $H$, and that $\tau$ sensitively depends on $\phi$, $H$ and the packing fraction. In particular, the ability of the oscillators to move and exchange neighbours can significantly decrease $\tau$. In contrast, model $\cal J$ predicts suppression of the synchronized state in subpopulation $\cal A$  and emergence of the coherent attractor in the ``contrarians'' subpopulation $\cal B$ for strong and weak cross-population coupling, respectively.   

\end{abstract}

\maketitle  

\section{Introduction}

Synchronization phenomena has been studied for a long time and is ubiquitous in numerous natural and artificial systems, for example,  in sociology, when people clap their hands together \citep{nda2000}, in ecology with  the synchronization of frog choirs \citep{aihara2008} or firefly blinking \citep{Ramrezvila2018}, in neuroscience with the periodic spiking of autonomous pacemaker neurons \citep{Plenz:1999,surmeier:2005} and neural synchronization \citep{GuevaraErra2017}, in physiology with biological rhythms \citep{Glass2001}, in active matter with the bacteria self-propulsion \citep{nIgoshin2001}, in nanotechnology with the synchronization of mechanical nanodevices \citep{antonio2015}, in condensed matter physics with spin Hall nano-oscillators \citep{Awad2016} etc.

\par Synchronization can be thought about as the {\it "adjustment of rhythms of oscillating objects due to their weak interaction"} \citep{arkadypikovsky2003}. Oscillating objects which have been most studies are so-called self-sustained oscillators, i.e., oscillators which can sustain their natural rhythm due to some internal energy source and that are normally stable to small perturbations, returning to their original rhythm when left by themselves. These objects most frequently are modeled as limit-cycle oscillators whose behavior is governed by autonomous non-linear differential equations \cite{barrat_barthelemy_vespignani_2008}. Depending on the topology of the interactions network of the oscillating objects as well as their physical nature, various types of synchronization are possible such as complete \cite{pecora:1990} and generalized \cite{rulkov:1995} synchronization. In this study we deal with a weaker form of synchronization, so-called phase synchronization \cite{kuramoto75,Rosenblum:1996} which can be realized in a system of coupled phase oscillators (their amplitude is irrelevant in this case) and described by the celebrated Kuramoto model  \cite{kuramoto75}.

A basic understanding of the general conditions for the existence and stability of the synchronized state of coupled oscillators is gained by applying the Master Stability Function formalism, which was developed by Pecora, Carroll and Barahona \cite{pecora:1998,barahona:2002,pecora:2013}. This method addresses a linear stability of the synchronized state of a generic system of linearly and symmetrically coupled dynamical objects, and allows to formulate a general criterion for the synchronizability of the network links between the dynamical objects [forming the network nodes] independently of their specific physical nature \cite{Arenas2008}. For a given nature of the objects, the criterion sets a constraint on the eigenvalues of the Laplacian matrix of the coupling network in order to exhibit synchronized behavior \cite{barahona:2002}.

Besides considering oscillators with fixed interaction networks, researchers investigated synchronization in time-varying network topologies \citep{Skufca2004,Stilwell:2006} meant to account for the effects of motility on coherent behavior of autonomous objects, for example in swarming. It was found  that networks with instantaneously disconnected topologies do allow for the synchronization of oscillators, provided the topology varies in time in a certain way \cite{Stilwell:2006}. A dynamic network topology realized through the motion of locally coupled mobile agents was considered in refs.~\cite{frasca:2008,peruani:2010,Fujiwara2011,uriu:2014,levis17}. A main common conclusion of these studies is that the mobility of the agents enhances synchronization. However, much richer behavior can be obtained when the interactions between the agents in the real space are included, e.g., the synchronization time exhibit a non-monotonic behavior as a function of the mobility of agents with excluded  volume interaction \cite{levis17}.

In numerous instances however synchronization is undesired \cite{louzada2012,Lameu:2016} as it can result in system malfunctioning \cite{louzada2012,Lameu:2016}. Examples include certain dysfunctions of the nervous system, such as epilepsy, Alzheimer and Parkinson diseases, which are thought to be caused by excessive neural synchronization  \cite{UHLHAAS2006,GuevaraErra2017,surmeier:2005,Kringelbach2007,Deuschl2006}, or dysfunctions in a computer network due to an unintended synchronization of the network routers  \cite{Floyd:1994}. Researchers addressed the issue of suppression of an undesired synchronization of phase oscillators in the context of the Kuramoto model with non-identical interactions \cite{zanette2005,hong2011a,hong2011b,louzada2012,mirchev2014,ratas2016}.
Thus, by introducing to the system a given number of ``contrarian'' oscillators, i.e., the oscillators which tend to dephase with all the other oscillators in the system, it is possible to completely suppress the global synchronization \cite{zanette2005}, or to change the nature of the synchronization transition from a first-order to second-order type \cite{zhang:2016}. Furthermore, by coupling the phase and spatial dynamics of oscillators it is possible to achieve very rich spatiotemporal behavior such as particle segregation according to their phases and phase waves, as reported recently in \citep{OKeeffe2017}.

Here we consider systems of phase oscillators which perform Brownian motion in a two dimensional box with periodic boundary conditions. Regarding the oscillator phase dynamics, we subdivide the system into two subpopulations, which we name subpopulation $\cal A$ and $\cal B$, respectively, and describe the dynamics of the oscillator phases in terms of a generalized Kuramoto model with local coupling of the oscillators [related to a finite range of interactions in the real space] and with a three-valued coupling constant, where each of the values describes the two  intra- and  one cross-population phase interactions. The main goal of this study is to understand how the synchronization within a target subpopulation, say of type $\cal A$, can be influenced by introducing ``controlling agents'' of type $\cal B$. In particular, we are interested in identifying under which conditions the synchronization in the target subpopulation can be significantly suppressed or eliminated completely. 

The rest of the paper is organized as follows: in the next section we introduce a model for Brownian dynamics of the oscillators in the real space and two different models for the phase dynamics. We also define synchronization order parameters and discuss some analytical results of the continuum mean-field approximation for the standard Kuramoto model. In section \ref{results} we discuss the simulation results emphasizing qualitative differences between the two models. In the last section we summarize our results and discuss the potential extension of the present work.

\section{Models: dynamics of the oscillators in the real and phase spaces}

\subsection{Brownian motion of oscillators with excluded volume}

We consider $N$ oscillators of mass $m$ in a two-dimensional (2D) box of size $L \times L$ and assume periodic boundary conditions in both spacial directions. The time-dependent position of the center of mass of oscillator $i$ is denoted by ${\bm r}_i(t)\equiv(x_i(t),y_i(t))^T$, where $i=1,..,N$ and $t$ stands for time. Without loss of generality we assume that the oscillators are dispersed in some fluid environment and that their mutual repulsion is described by interaction potential $U(\rvec_1,...,{\bm r}_N)$. The time evolution of  ${\bm r}_i(t)$ is described by a system of coupled Langevin equations:
\begin{equation}
m\frac{d^2 {\bm r}_i }{dt^2} =  -\gamma \frac{d {\bm r}_i }{dt} - \nabla_i U(\rvec_1,...,{\bm r}_N) + \sqrt{2 \gamma \kB T }\, {\bm \xi_i}(t),
\label{eq:langevin}
\end{equation}
where the first term on the r.h.s. describes the friction force (due to, e.g., interaction with the surrounding ambient fluid) with the friction coefficient $\gamma$,  $\kB$ denotes the Boltzmann constant and $T$ is the absolute temperature of the ambient fluid. $\gamma$ is related to the diffusion coefficient $D$ by the fluctuation-dissipation theorem $D= \kB T/\gamma$. The second term gives the force on oscillator $i$ originating from the repulsive interactions with the other oscillators characterized by the interaction potential $U$, $\nabla_i$ stands for the gradient operator acting upon $\rvec_i$. Thermal fluctuations in surroundings are accounted for by a random force ${\bm \xi_i}(t)$, which is a zero-mean delta-correlated Gaussian process, i.e., $\langle {\bm \xi_i}(t) \rangle=0$, $\langle {\bm \xi}_i(t_1){\bm \xi}_j^T(t_2) \rangle=\delta_{ij} {\bm 1} \delta(t_1-t_2)$ with $ {\bm 1}$ the identity matrix, $\delta$ the Dirac delta function, and $\langle ... \rangle$ denoting an ensemble average. We approximate $U$ as a pairwise sum of Lennard-Jones truncated and shifted potential
\begin{equation}    
    U(\rvec_1,...,\rvec_N) = \begin{cases} 4\epsilon\sum\limits_{i<j}^N{\left ( \left (\frac{\sigma}{r_{ij}} \right)^{12}- \left (\frac{\sigma}{r_{ij}}\right )^{6} \right )}, r_{ij} \leq \sqrt[6]{2}, \\ 0,  r_{ij} > \sqrt[6]{2}, \end{cases}
\label{eq:repulsive_Poential}
\end{equation}
 \noindent where $\epsilon$ defines the depth of the pair potential at its minimum corresponding to the particle separation $r_{ij} \equiv \lvert\lvert \rvec_i-\rvec_j\rvert \rvert=\sqrt[6]{2}$, and $\sigma$ is the effective diameter of the oscillators.
 
We have integrated Eqs.~(\ref{eq:langevin}) numerically by employing Large-scale Atomic/Molecular Massively Parallel Simulator (LAMMPS) package \citep{plimpton95} at constant volume $V=L^2$ and temperature $T=1.0 \epsilon/\kB$.  The molecular dynamics simulations were run with a time step $\Delta t = 0.01\sqrt{m\sigma ^2/\epsilon}$. The temperature was kept constant by using Langevin thermostat as described in \citep{Schneider1978} with the Langevin friction coefficient $\gamma = 1.0 \sqrt{m \epsilon/\sigma ^2}$  modeling a background implicit solvent. These values of $T$ and $\gamma$ set the  diffusion constant $D = 1.0 \sqrt{\sigma ^2 \epsilon/m}$.  

\subsection{Generalized Kuramoto model}

\subsubsection{The standard Kuramoto model and the synchronization order parameter}

 Yoshiki Kuramoto introduced his celebrated model in 1975 \citep{kuramoto75} in response to the earlier work on synchronization of biological rhythms by Winfree in 1967 \citep{winfree67}. The Kuramoto model can be solved exactly within certain limits, exhibiting a synchronization phase transition in systems of phase oscillators. This model is widely used to study synchronization phenomena in both dynamical and static networks of interacting oscillators \citep{Rodrigues2016,Schmidt2015}. For a system of $N_0$ all-to-all interacting phase oscillators, the dynamics of the phase $\theta_k(t)$ of oscillator $k$ is postulated in the following form 
\begin{equation}\label{eq:kuramoto_normal}
\frac{d \theta_k(t)}{dt} =\omega_k + \frac{K_0}{N_0}  \sum^{N_0}_{l=1}\sin(\theta_{l}(t) - \theta_{k}(t)), 
\end{equation}
\noindent where $\omega_k$ is the natural frequency of the $k$th oscillator when it is not interacting with any other oscillator and $K_0$ is a coupling constant. 
The synchronization often is analyzed in terms of a complex order parameter $z(t)$ defined as
\begin{equation}
\label{eq:kuramoto_parametroordem}
    z(t)=r(t)e^{i\Psi(t)}=\frac{1}{N_0}\sum_{k=1}^{N_0}{\mathrm e}^{i\theta_k(t)},
\end{equation}
\noindent with $r(t)$ quantifying the global oscillator coherence, and $\Psi(t)$ being the average phase; here $i=\sqrt{-1}$ is the imaginary unit. 
For a perfectly synchronized system $r(t) =1$, with $r(t)=0$ in the opposite limit of totally dephased oscillators. 
\par In the limit $N_0 \rightarrow \infty $ and for $\omega_k=0, k=1,..,N_0$ Ott and Antonsen derived the following mean-filed equation for $r(t)$ \citep{0806.0004}
\begin{equation}
\frac{d r (t)}{ d t}=\frac{K_0}{2}\left( r(t) - r(t)^3\right),
\label{eq:Ott_Antonsen_result}
\end{equation}
\noindent which for a given initial condition $r(0)$ has the solution 
\begin{equation}
    r(t) = \frac{1}{\sqrt{1+{\mathrm e}^{-K_0(t-t^*)}}},
    \label{eq:kuramoto_r(t)}
\end{equation}

\noindent with 
\begin{equation}
    t^* = \frac{\ln \left (\frac{1}{r(0)^2}-1 \right )}{K_0}.
    \label{eq:kuramoto_t_0}
\end{equation} 
\noindent Eq.~(\ref{eq:kuramoto_r(t)}) is the square root of the logistic function and their shapes are quit similar. After some initial transient regime, $K_0^{-1}$ defines a characteristic synchronization time scale on which $r(t)$  reaches its saturation value of $1$. Indeed, for $t\gg K_0^{-1}$  
 \begin{equation}
 r(t)\sim 1 -\frac{1}{2}{\mathrm e}^{-K_0(t-t^*)}.
 \label{eq:rt_assymptotics}
\end{equation}
$t^*$ marks the location of the inflection point of $r(t)$. Below we show, that in certain regimes the synchronization curves in our systems can be very accurately described by Eq.~(\ref{eq:kuramoto_r(t)}).

\begin{figure}[h!]
    \captionsetup{position=top,singlelinecheck=false,justification=raggedright}
    \subfloat[]{
    \includegraphics[height=0.49\linewidth,width=0.49\linewidth]{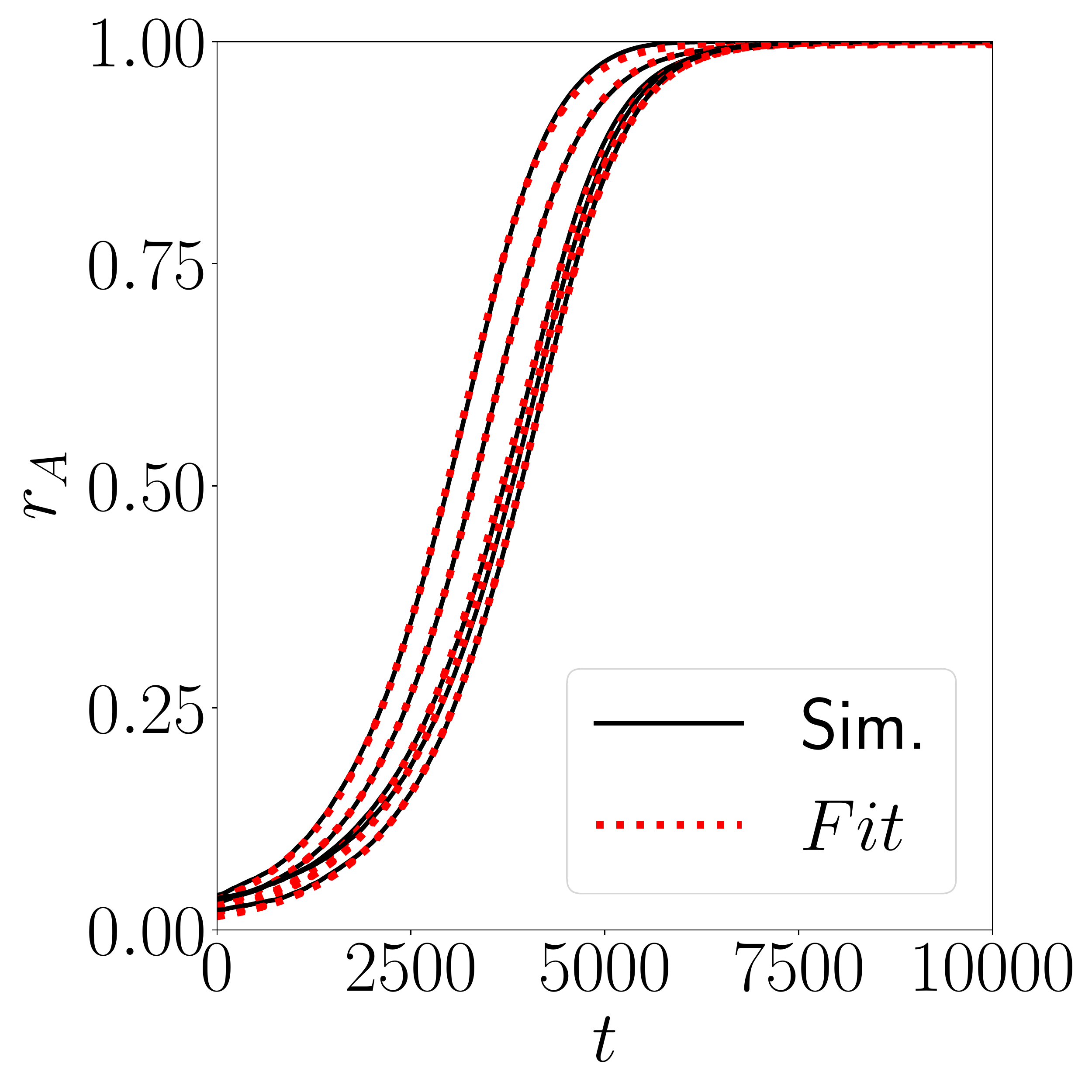}
    }
    \subfloat[]{
    \includegraphics[height=0.49\linewidth,width=0.49\linewidth]{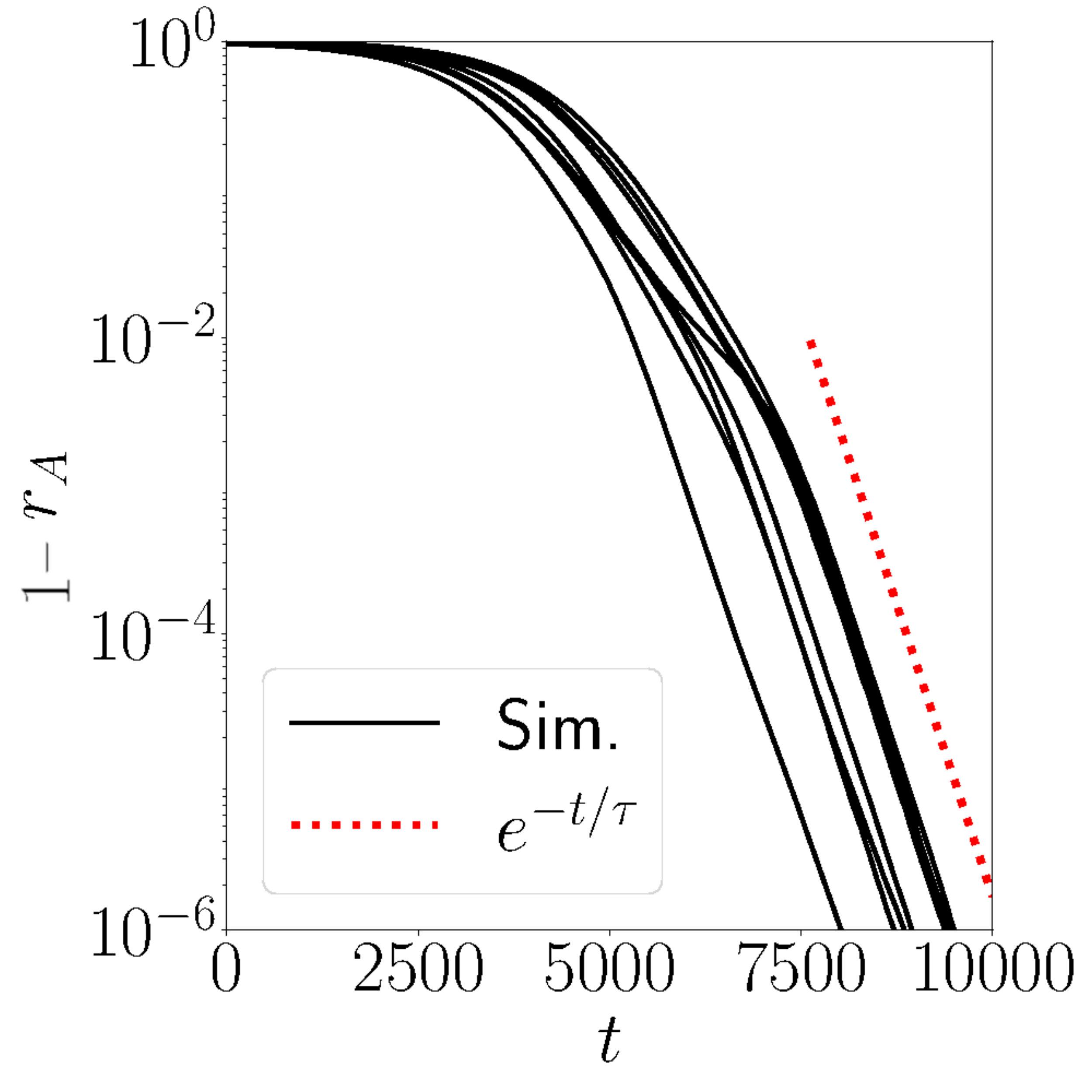}
    } \\
    \vspace{-1.10\baselineskip}
    \subfloat[]{
    \includegraphics[height=0.49\linewidth,width=0.49\linewidth]{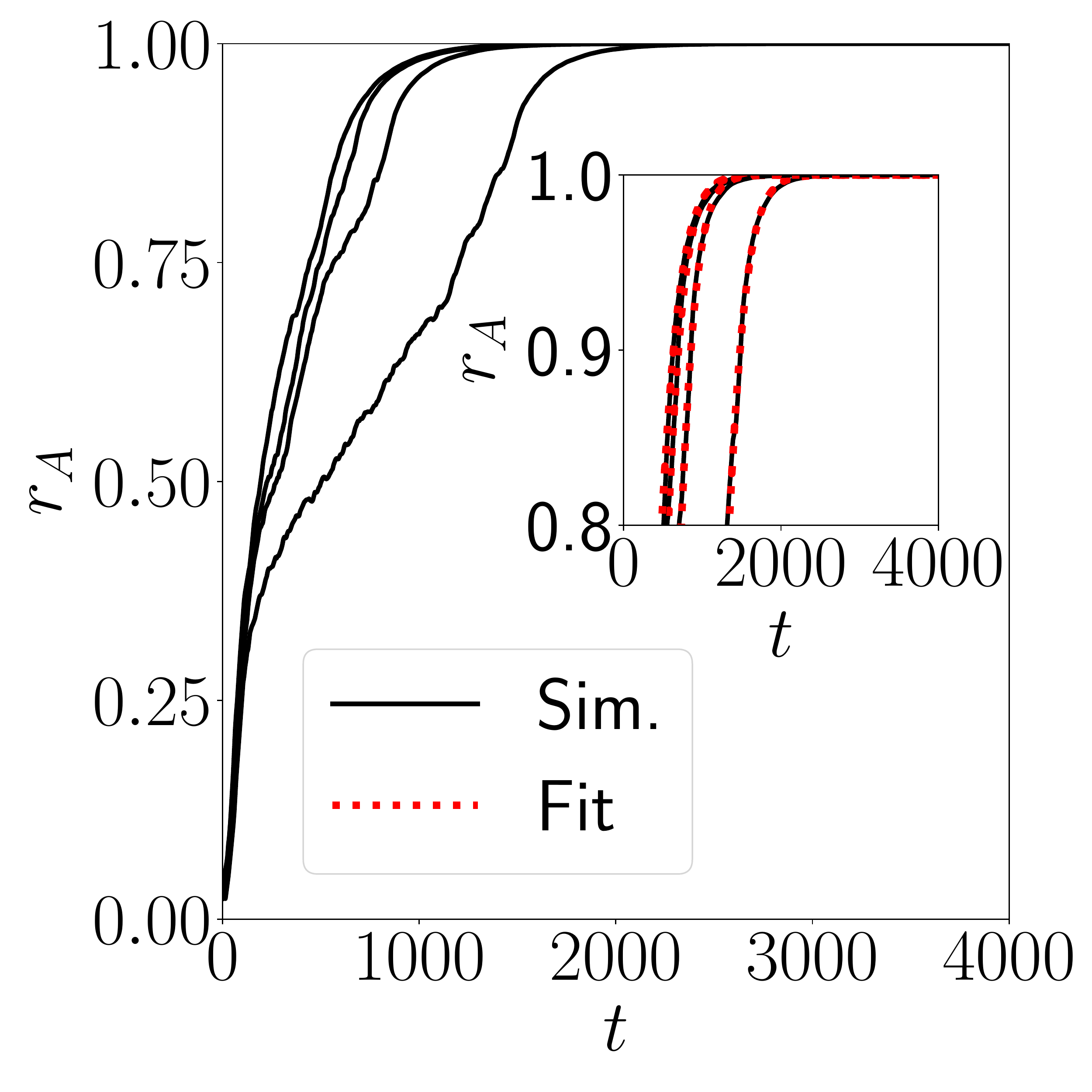}
    }
     \subfloat[]{
    \includegraphics[height=0.49\linewidth,width=0.49\linewidth]{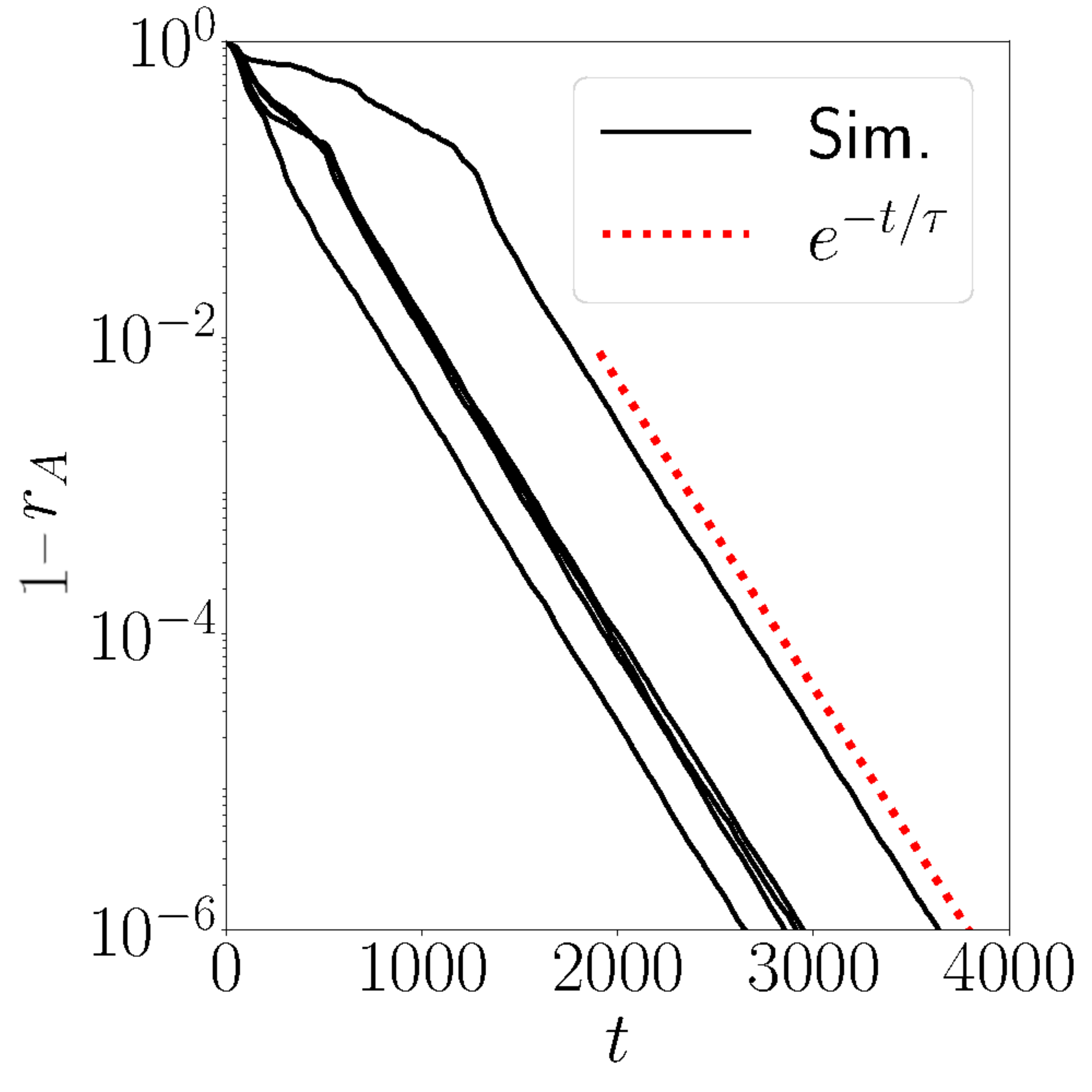}
    }
   \vspace{-0.5\baselineskip}
%    \captionsetup{position=top}
%\clearcaptionsetup[]
\centering
  \captionsetup{justification=centerlast,singlelinecheck=false}
      \caption{Panels (a) and (c) show $r_{\cal A}(t)$ for systems at $\zeta=1$ and $\zeta=100$, 
      respectively, other parameters are $N$ = 1800, $L=50\sigma$, $\phi = 0.5$, $K = 0.001\sqrt{\epsilon/m\sigma ^2}$. Different curves represent different system realizations, i.e., different initial $\theta_k^{\sigma}(t=0)$. Large fluctuations of $r_{\cal A}(t)$ for intermediate times in (c) are due to the annihilation dynamics of topological defects in the $\theta$ field, see Figs.~\ref{fig:binarysynchronizationpanel}(f) and (g). The red curves in (a) and in the inset of (c) represent fits to the mean-field model $r_{\cal A}(t) = 1/\sqrt{1+e^{-\kappa(t-t_0)}}$. In (c) the fitting was carried out only within the time range that corresponds to $r_{\cal A}\ge 0.8$. Panels (b) and (d) highlights the late time, $\kappa t \ll 1$, behavior of $1-r_{\cal A}(t)$ with exponential decay $1 - r_{\cal A}(t) \sim e^{-t/\tau}$. We find that $\kappa^{-1}\ne \tau$. Time is given in units of $\sqrt{m\sigma ^2 /\epsilon}$.
      \label{fig:fitpanel}
    }
    
\end{figure}

\subsubsection{Models for binary mixtures of phase oscillators }

%\par The motion of our oscillators will be given by the Langevin dynamics described in the previous chapter, equation (\ref{eq:langevinmotion}). We will be using LJ units as described in equations (\ref{iniciounidades}-\ref{fimunidades}).
We subdivide the system of $N$ phase oscillators in $N_{\cal A}$ oscillators of type $\cal A$, whose phases we denote by  $\theta_{k}^{\cal A}$, and in $N_{\cal B}$ oscillators of type $\cal B$, whose phases we denote by  $\theta_{k}^{\cal B}$, and $N=N_{\cal A}+N_{\cal B}$. 
We consider  that $\cal A-$oscillators form the target subpopulation whose synchronization behavior will be manipulated through the addition of the control oscillators of type $\cal B$.  
The governing equations of our model are as follows 
%\begin{eqnarray}
%\label{eq:modkuramotoa}
%\frac{{d\theta}_{k \in A}}{dt} =\omega_k + \frac{1}{N_{R}} \Biggl [ K_{AA}\mathlarger{\mathlarger{\sum}}_{l\in A,r_{kl}<R}\sin \Bigl (\theta_{l} - \theta_{k} \Bigr) + \nonumber \\ 
%K_{AB}\mathlarger{\mathlarger{\sum}}_{l\in B,r_{kl}<R}\sin \Bigl (\theta_{l} - \theta_{k}\Bigr )\Biggr ],
%\end{eqnarray}
%
%\begin{eqnarray}
%\label{eq:modkuramotob}
%\dot{\theta}_{i \in B} =\omega_i + \frac{1}{n_{i}} ( \sum^{n_i}_{j\in A}H\sin(\theta_{j} - \theta_{i})+\sum^{n_i}_{j\in B}K\sin(\theta_{j} - \theta_{i})),
%\frac{{d\theta}_{k \in B}}{dt} =\omega_k + \frac{1}{N_{R}} \Biggl [K_{BA} \mathlarger{\mathlarger{\sum}}_{l\in A,r_{kl}<R}\sin \Bigl (\theta_{l} - \theta_{k} \Bigr) + \nonumber \\ 
%K_{BB}\mathlarger{\mathlarger{\sum}}_{l\in B,r_{kl}<R}\sin \Bigl (\theta_{l} - \theta_{k}\Bigr )\Biggr ],
%\end{eqnarray}
%
\begin{align}
\begin{split}
\label{eq:modkuramoto}
\frac{{d\theta}_{k}^{\sigma}}{dt}  &= \omega_k^{\sigma} \\
 &+  \frac{1}{N_{R}} \mathlarger{\mathlarger{\sum}}_{\sigma^{\prime} 
  ={\cal A,B}} K_{\sigma\sigma^{\prime}}\mathlarger{\mathlarger{\sum}}_{l=1,r_{kl}<R}^{N_{\sigma^{\prime}}}\sin \Bigl (\theta_{l}^{\sigma^{\prime}} - \theta_{k}^{\sigma} \Bigr),
\end{split}
\end{align}
\noindent where $N_R$ is the instantaneous number of the neighbours of oscillator $k$ determined by the interaction range $R$, i.e., is the number of oscillators in a disk with the radius $R$ and centered at $\rvec_k$.
Throughout the whole study we set $R = 3 \sigma$ and assume for simplicity $\omega_k = 0$ for every $k$. We consider two specific choices of the $2\times2$ matrix $K_{\sigma\sigma^{\prime}}$  of the coupling constants, which we designate as model $\cal I$ and model $\cal J$. In model $\cal I$ we use a symmetric form
\begin{equation}
  \begin{bmatrix}
    K_{\cal AA} & K_{\cal AB}  \\
    K_{\cal BA} & K_{\cal BB}
    \end{bmatrix}  =\begin{bmatrix}
    K & H  \\
    H & K
    \end{bmatrix},
    \label{eq:modelI}
\end{equation}
where $K>0$ and $H<0$ are the coupling constants for alike and unlike oscillators, respectively. Positive $K$ drive synchronization within a given subpopulation, while the negative $H$ forces distinct oscillators to proceed out of phase. 

\begin{figure*}
    \captionsetup{position=top,justification=raggedright,singlelinecheck=false}
    \centering
    \subfloat[]{
    \centering
    \includegraphics[width=0.22\linewidth]{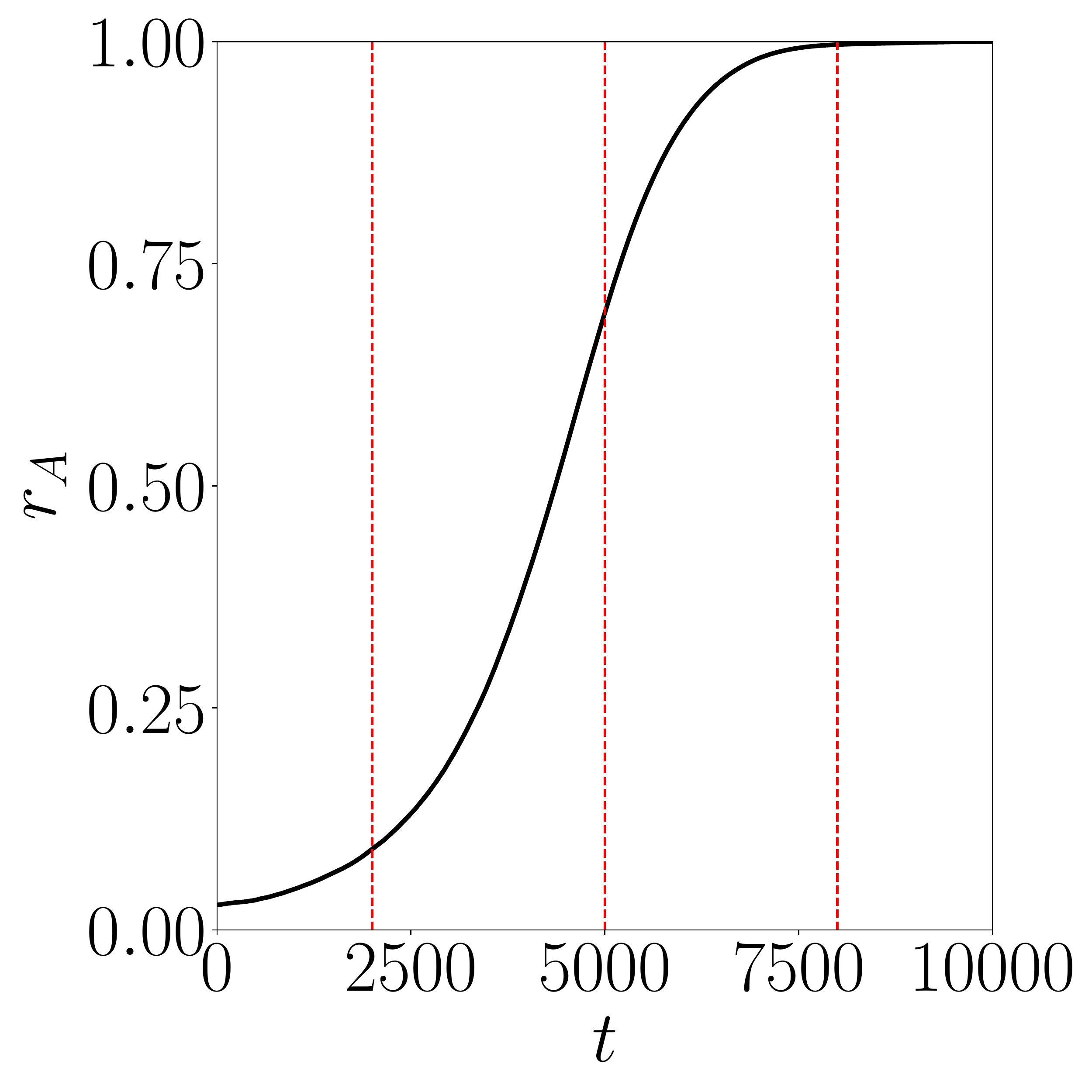}
    }
     \subfloat[]{
    \centering
    \includegraphics[width=0.22\linewidth]{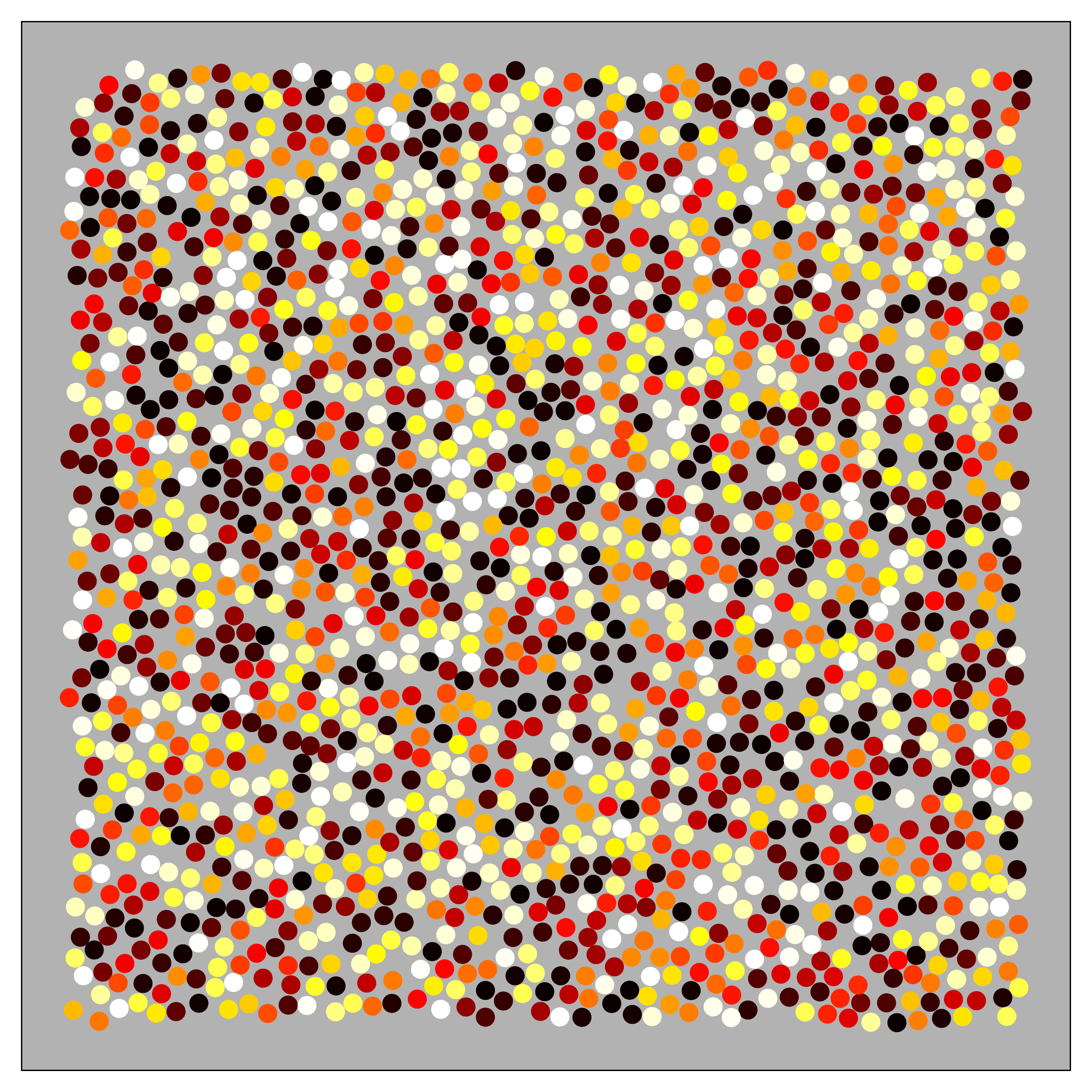}
    } 
    \centering
    \subfloat[]{
    \centering
    \includegraphics[width=0.22\linewidth]{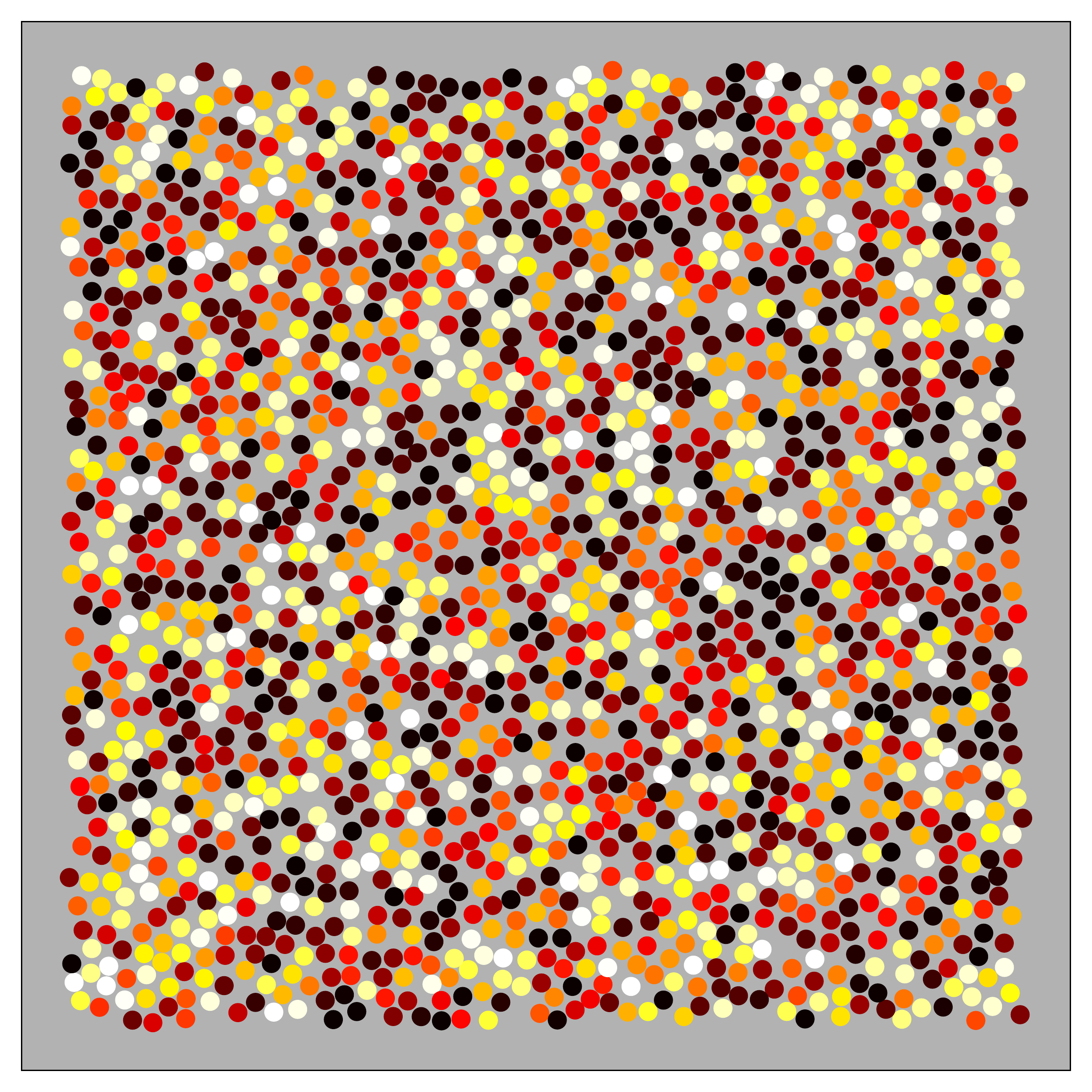}
    }
     \subfloat[]{
    \centering
    \includegraphics[width=0.22\linewidth]{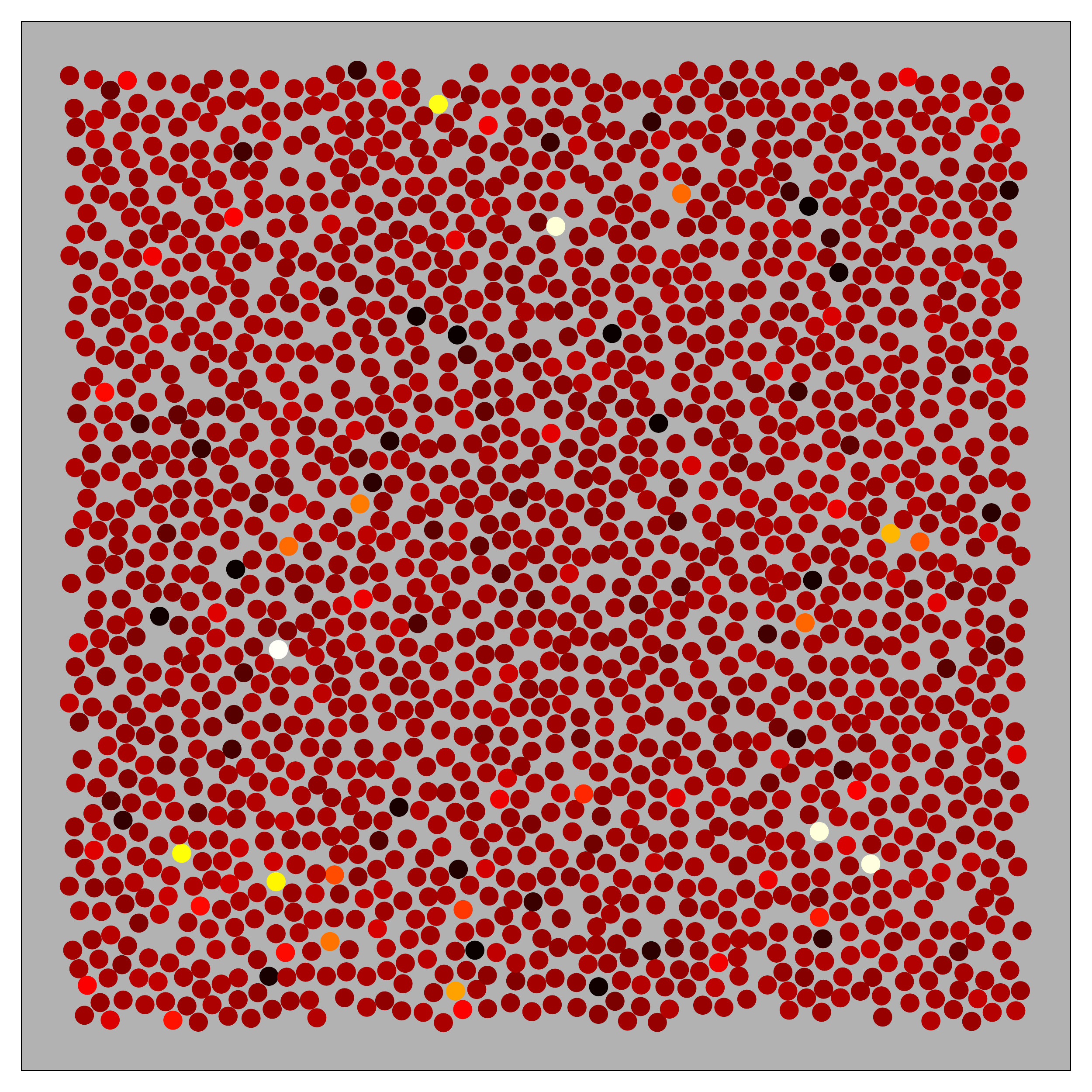}
    } \\
    \centering
    \subfloat[]{
    \centering
    \includegraphics[width=0.22\linewidth]{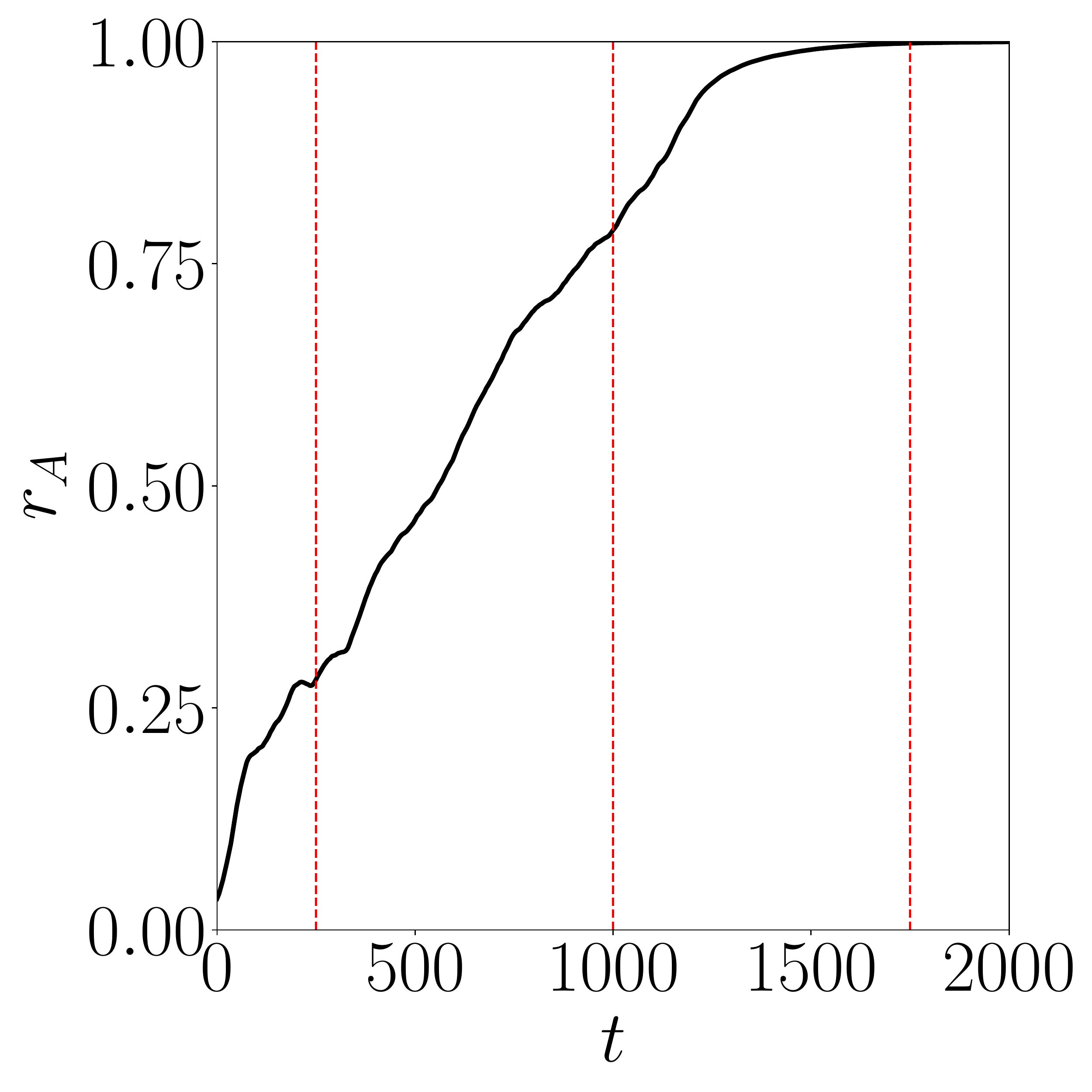}
    }
     \subfloat[]{
    \centering
    \includegraphics[width=0.22\linewidth]{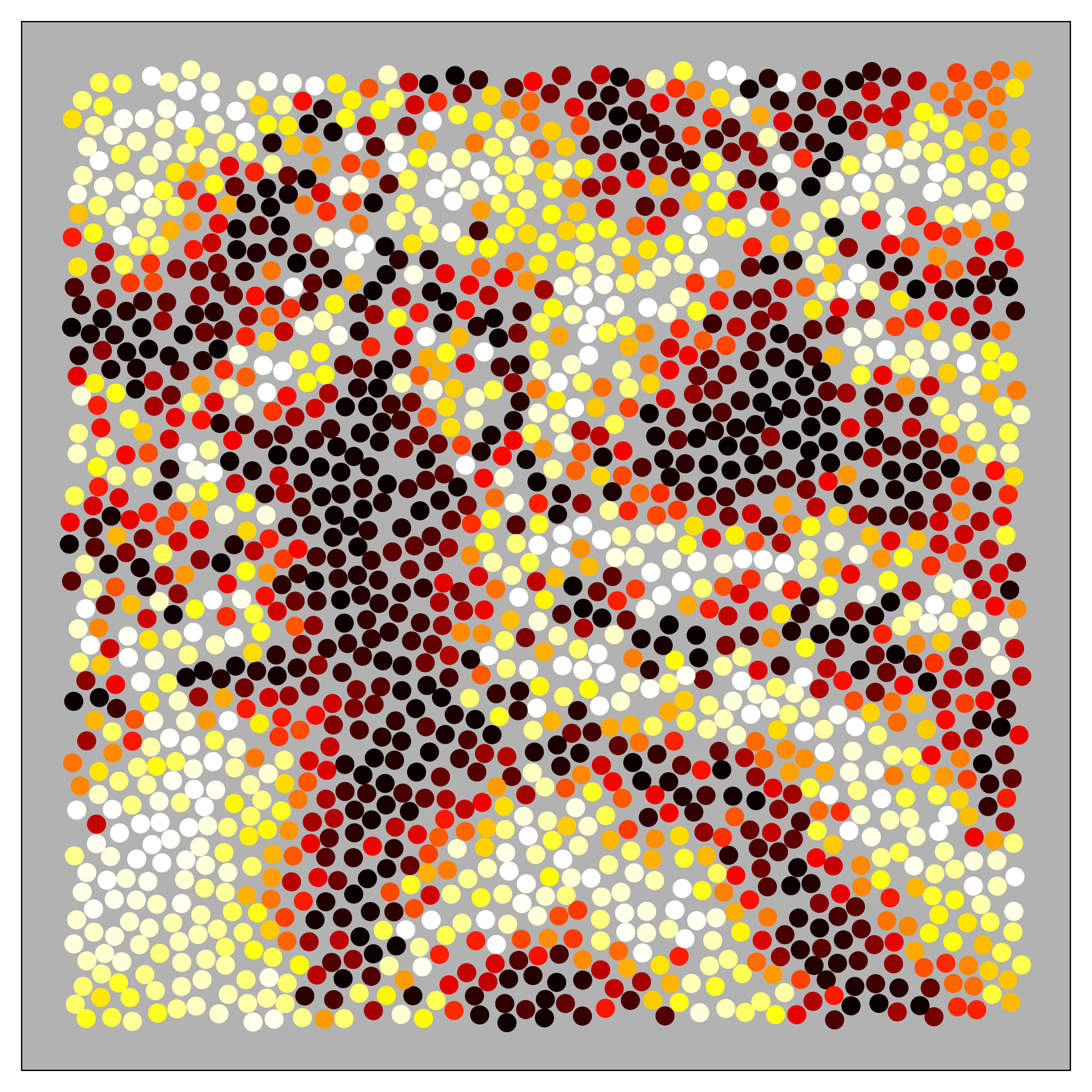}
    } 
    \centering
    \subfloat[]{
    \centering
    \includegraphics[width=0.22\linewidth]{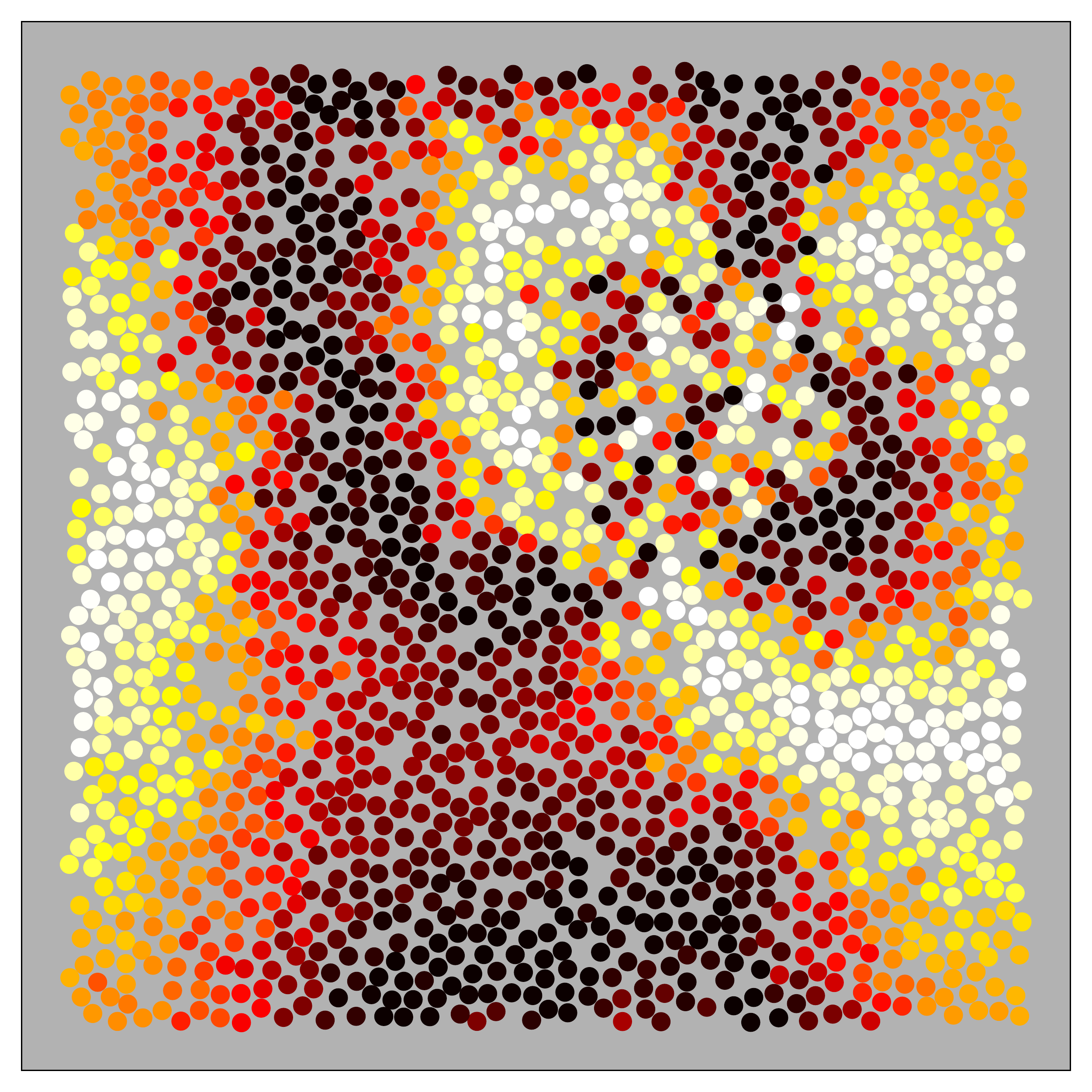}
    }
    \subfloat[]{
    \centering
    \includegraphics[width=0.22\linewidth]{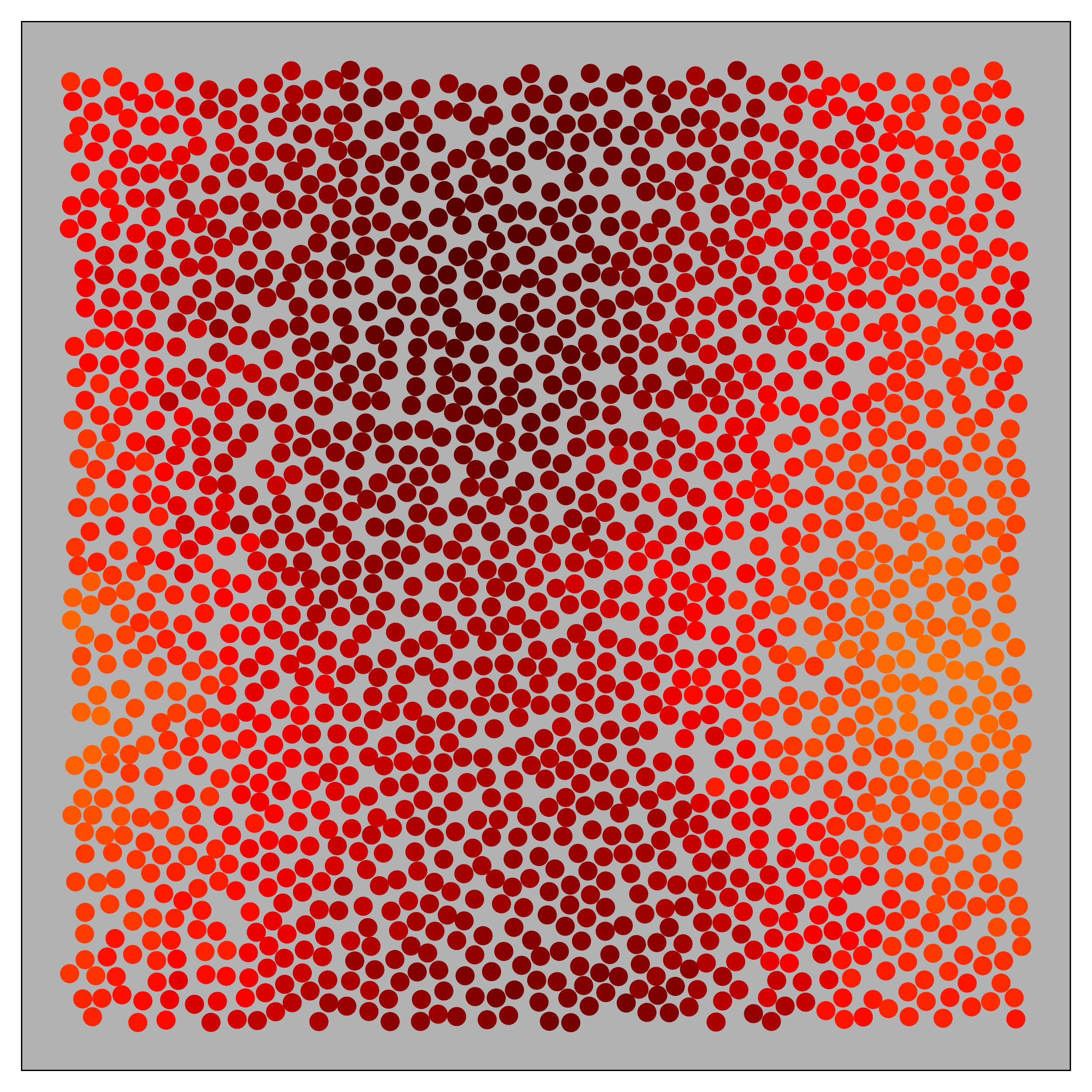}
    }\\
     \subfloat{
    \centering
    \includegraphics[height=0.10\linewidth]{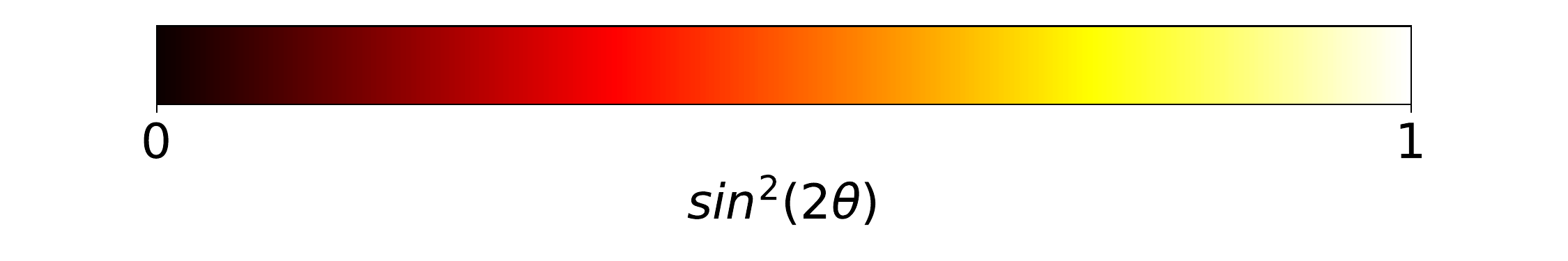}
    }
    \vspace{-0.5\baselineskip}
    \captionsetup{justification=centerlast,singlelinecheck=false,singlelinecheck=true}
    \caption{(a) $r_{\cal A}(t)$,  (b)-(d) the instantaneous oscillator configurations at times matching those marked by vertical lines in (a). The represented system corresponds to fast switching regime, with $N$ = 1800, $L=50\sigma$, $\phi = 0.5$, $K = 0.001\sqrt{\epsilon/m\sigma ^2}$ and $\zeta = 1$. Panels (e)-(h), the same as above but at $\zeta = 100$. Both $\cal A-$ and $\cal B-$oscillators are shown and colored according to $\mathrm {sin}^2(\theta_k^{\sigma})$. We choose this color scheme in order to emphasize the presence of the topological defects. Both systems reach globally synchronized configurations, $r \approx 1$.
    }
  \label{fig:binarysynchronizationpanel}
\end{figure*}

In model $\cal J$ we assume that each $\cal B-$oscillator has negative coupling to all the other $N$ oscillators in the system and set
\begin{equation}
  \begin{bmatrix}
    K_{\cal AA} & K_{\cal AB}  \\
    K_{\cal BA} & K_{\cal BB}
    \end{bmatrix}  =\begin{bmatrix}
    K & H  \\
    H & H
    \end{bmatrix}.
    \label{eq:modelII}
\end{equation}
%
%These equations can be integrated in the LAMMPS framework by assigning to every oscillator a new intrinsic degree of freedom, the phase, and integrating equation (\ref{eq:kuramoto_normal}) using the Euler method.
We define composition $\phi$ of the binary mixture as $\phi = N_{\cal B}/(N_{\cal A}+N_{\cal B})$ which controls the average number of repulsive pairwise contacts in the system. We also introduce packing fraction $\eta=N\pi \sigma^2/4L^2$, and parameter $\zeta \equiv |H|/K$ quantifying the interaction asymmetry of mixture. Below we report the effects of these three parameters on the synchronization of subpopulation $\cal A$ which is quantified by the order parameter
\begin{equation}
\label{eq:orderparameter}
r_{\cal A}(t) = Re\left(\frac{1}{N_\sigma}\mathlarger{\sum}_{k =1}^{N_{\cal A}}{\mathrm e}^{i\theta_k(t)}\right).
\end{equation}
The order parameter for subpopulation $\cal B$ can be defined in a similar way.

\begin{figure}[h!]
    \captionsetup{position=top,justification=raggedright,singlelinecheck=false}
    \centering
    \subfloat[]{
    \centering
    \includegraphics[width=0.48\linewidth]{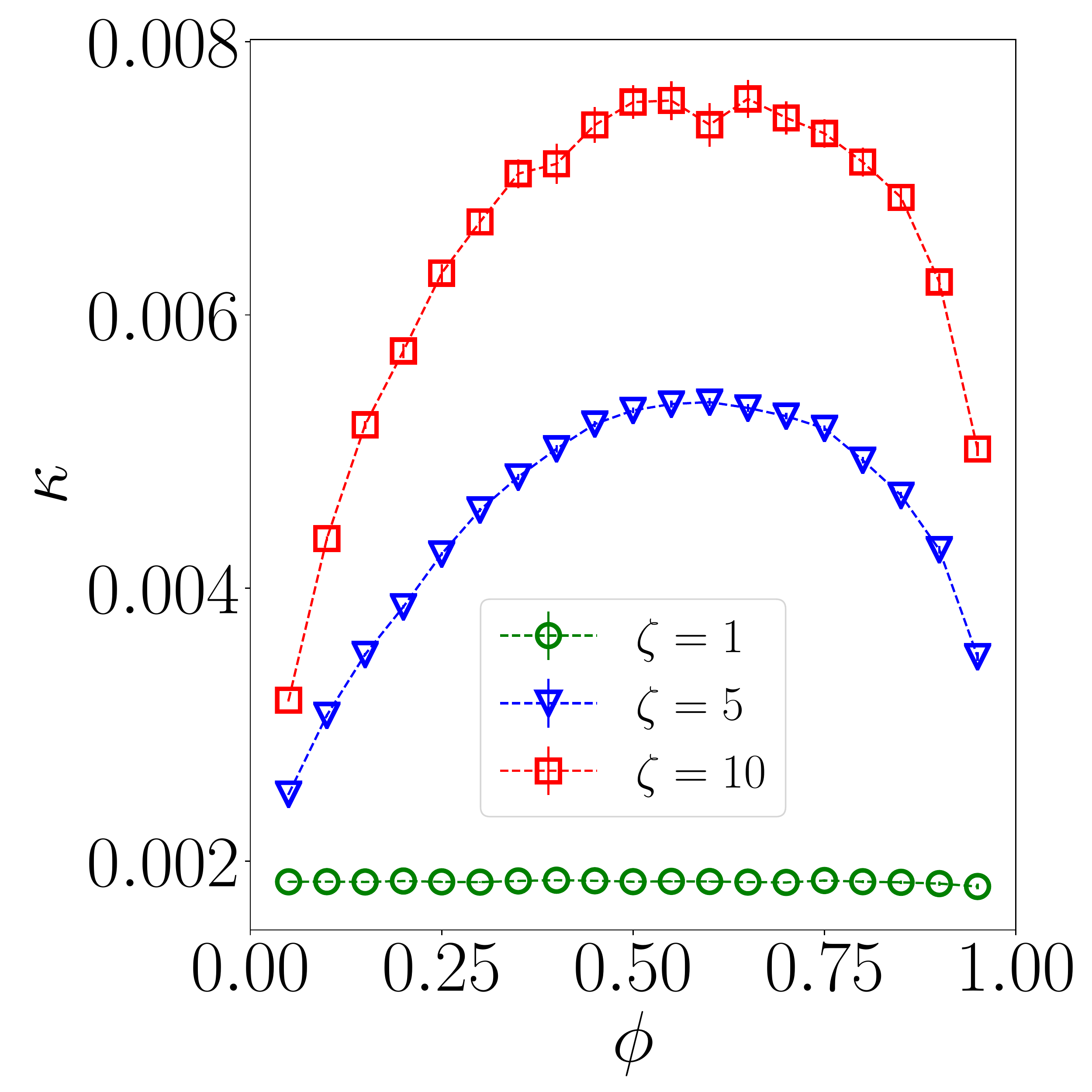}
    }
     \subfloat[]{
    \centering
    \includegraphics[width=0.48\linewidth]{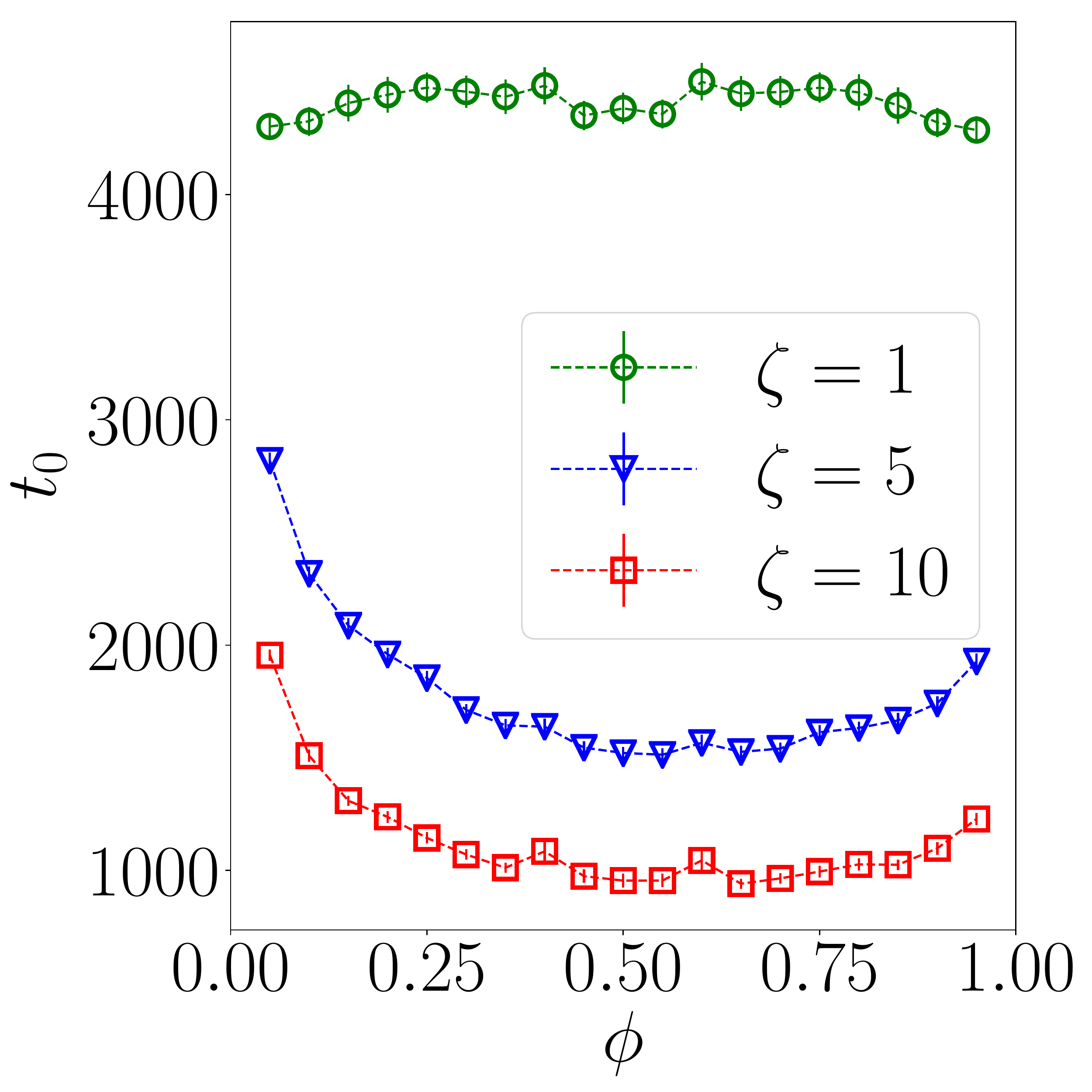}
    } 
    \vspace{-0.5\baselineskip}
    \captionsetup{justification=centerlast,singlelinecheck=false}
    \caption{(a) $\kappa$, (b) $t_0$ as functions of the composition $\phi$ of the binary mixture for several values of $\zeta$. $\kappa$ and $t_0$ are plotted in units of $\K$ and $\tunit$, respectively. In all the cases, $K=0.001\K$, $N = 1800$ and $L=50\sigma$. The results have been obtained after averaging over 100 independent runs.}
    \label{fig:KappaT0phi}
\end{figure}

We integrate equations (\ref{eq:modkuramoto}) numerically by using the Euler method with the same time step as is used to integrate Eqs.~(\ref{eq:langevin}), i.e., $\Delta t=0.01\sqrt{m\sigma ^2/\epsilon}$. 
Specifically, we have implemented this integration in a fictitious LAMMPS pair style function, which is then superimposed with the particle pair potential in the real space. The Kuramoto dynamics, as given by Eqs.~(\ref{eq:modkuramoto}), is switched on only after the system of soft disks, whose dynamics in the real space is governed by Eqs.~(\ref{eq:langevin}), reaches thermal equilibrium. Initial phases $\theta_k(t=0),k=1,..,N$ are set to random values drawn from uniform distribution on $[0,2 \pi)$.

\section{Results}
\label{results}
\subsection{Model $\cal I$}

\begin{figure}[h!]
  \captionsetup{position=top,justification=raggedright,singlelinecheck=false}
 
    \subfloat[]{
    \centering
    \includegraphics[height=0.49\linewidth,width=0.49\linewidth]{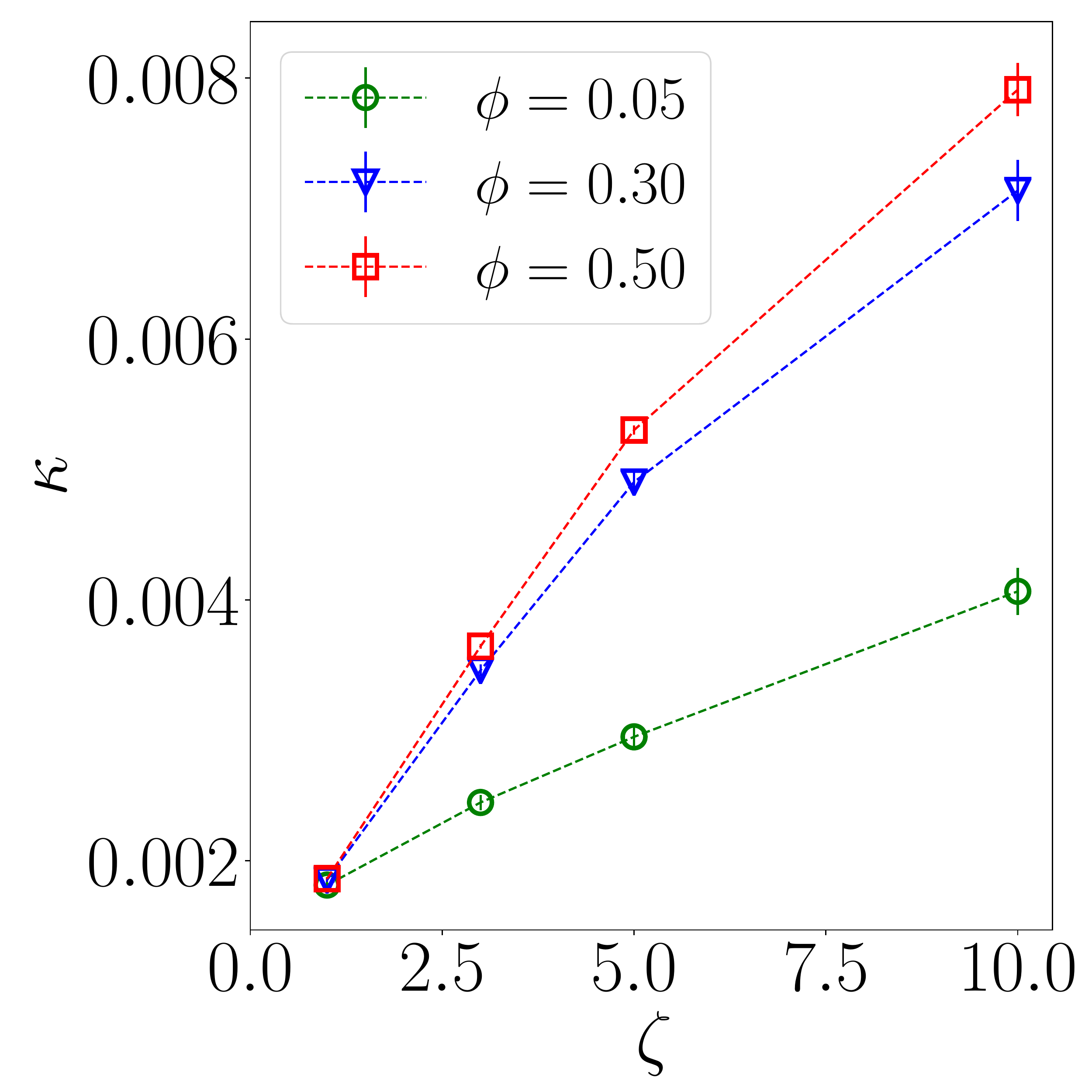}
    }
     \subfloat[]{
    \centering
    \includegraphics[width=0.49\linewidth]{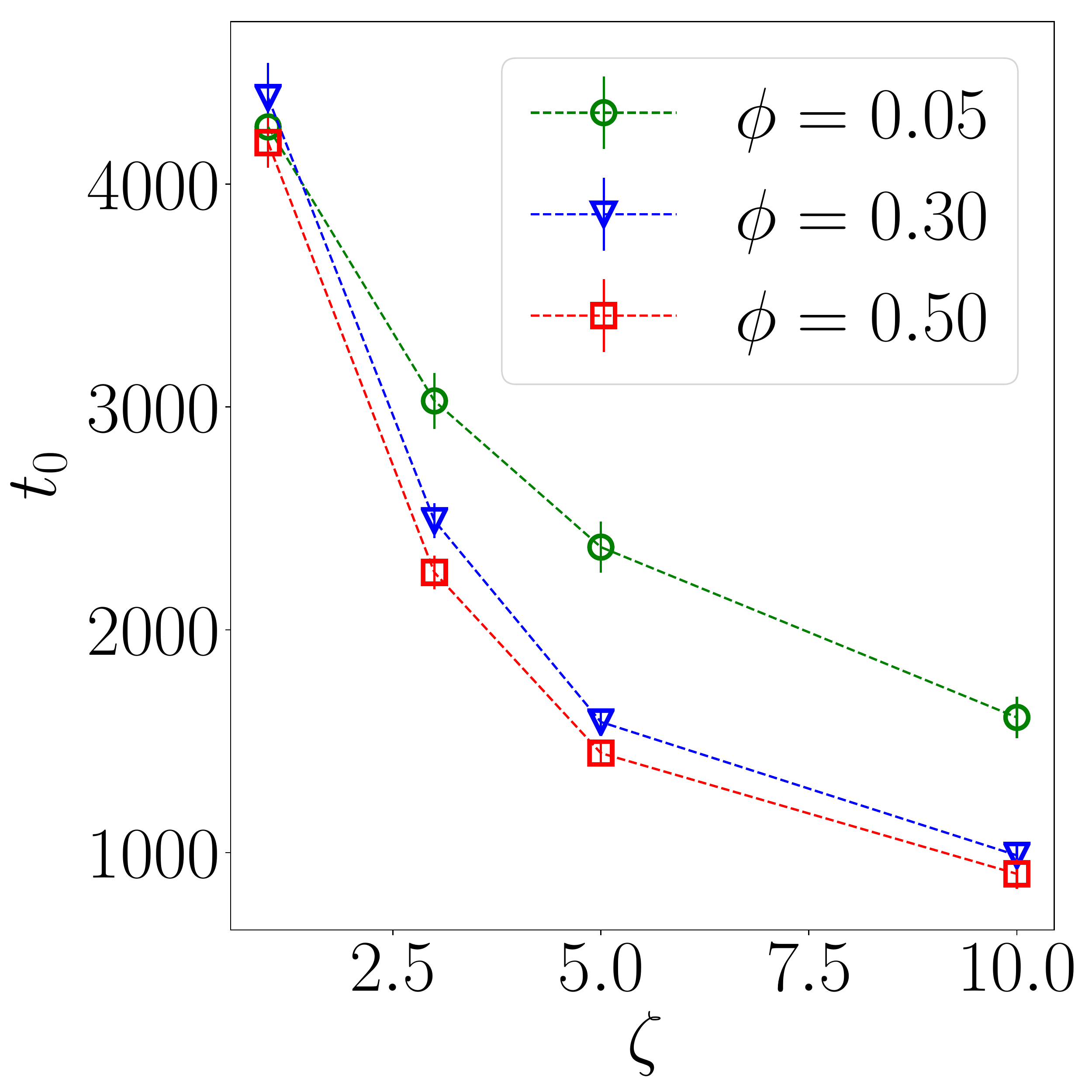}
    }
 %  \captionsetup{justification=center}
 \vspace{-0.5\baselineskip}
    \caption{ (a) $\kappa$ and (b) $t_0$ as functions of $\zeta$ for several values of the mixture composition $\phi$. $\kappa$ is plotted in units of $\K$, and $t_0$ in units of $\tunit$.  In all the case $K=0.001\K$, $N = 1800$ and $L=50\sigma$. The results are obtained after averaging over 100 independent runs.
  }\label{fig:KappaT0H}
\end{figure}

We introduce the time scale of local synchronization $t_{\theta}=\min(K^{-1},{H}^{-1})$, and the second timescale $t_r$ associated with the rate of change of the network topology due to the motion of oscillators $t_r = R^2/D$. $t_r$ characterizes an average time that an oscillator requires to diffuse over a distance of the interaction range $R$. Equivalently $t_r$ can be thought of as an average duration of the interaction between the phases of two oscillators before they diffuse apart beyond the interaction range. In our case $t_r=9\sqrt{m\sigma ^2 /\epsilon}$. The limit $t_{\theta}\gg t_r$ corresponds to so-called fast switching [FS] regime in the language of dynamic networks, when the network links change much faster as compared to the local phase reorientation. Previous studies of the synchronization of identical mobile oscillators [locally coupled] revealed that in the FS regime one can carry out time averaging of the original time-dependent adjacency matrix, which renders a mapping of the phase dynamics of mobile oscillators onto that of an effective all-to-all static Kuramoto model \cite{frasca:2008,Fujiwara2011}. Therefore, in the FS limit the model can be treated at the mean-field level \cite{levis17}.

\begin{figure*}
    \centering
    \captionsetup{position=top,justification=raggedright,singlelinecheck=false}
    
    \subfloat[]{
    \centering
    \includegraphics[width=0.22\linewidth]{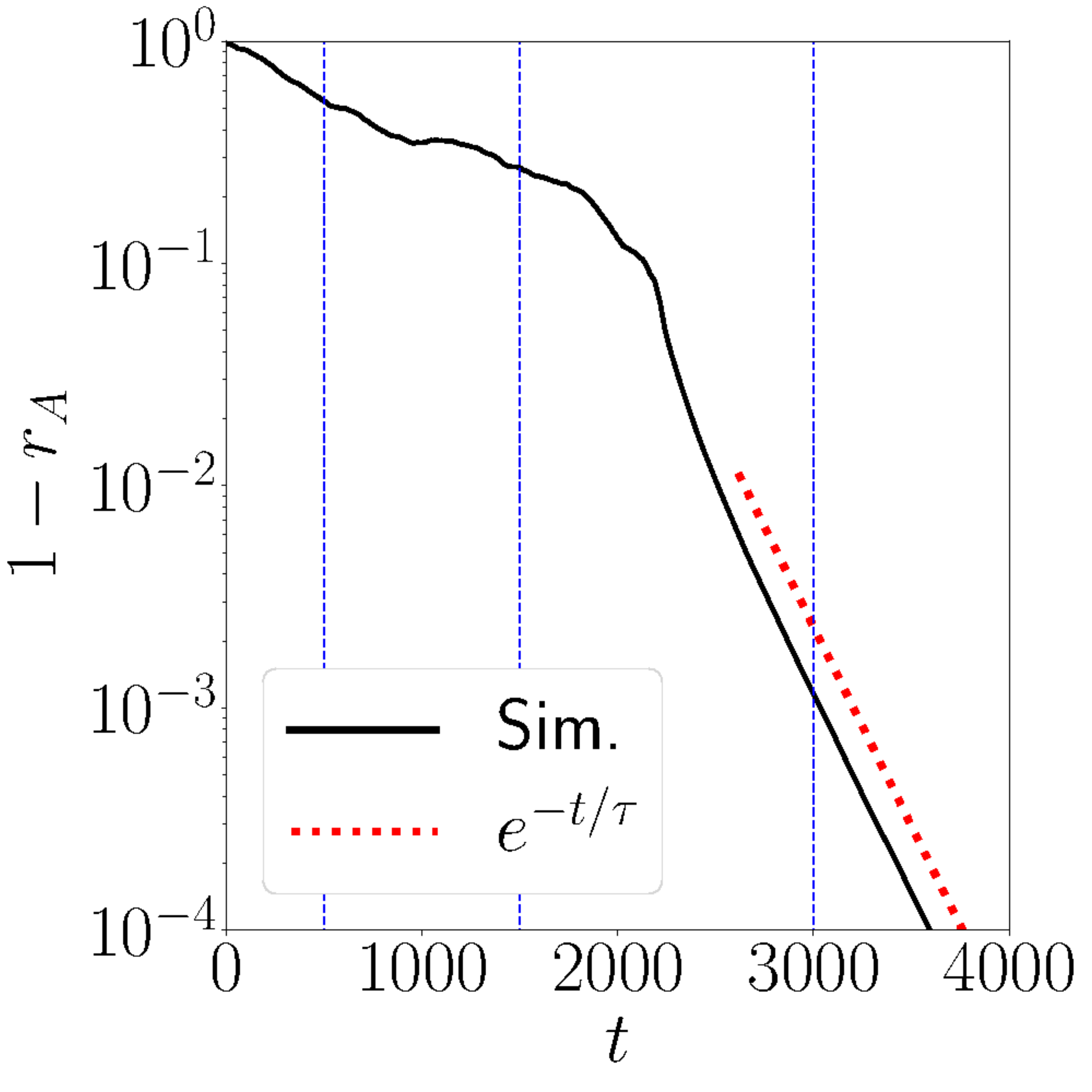}
    }
     \subfloat[]{
    \centering
    \includegraphics[width=0.22\linewidth]{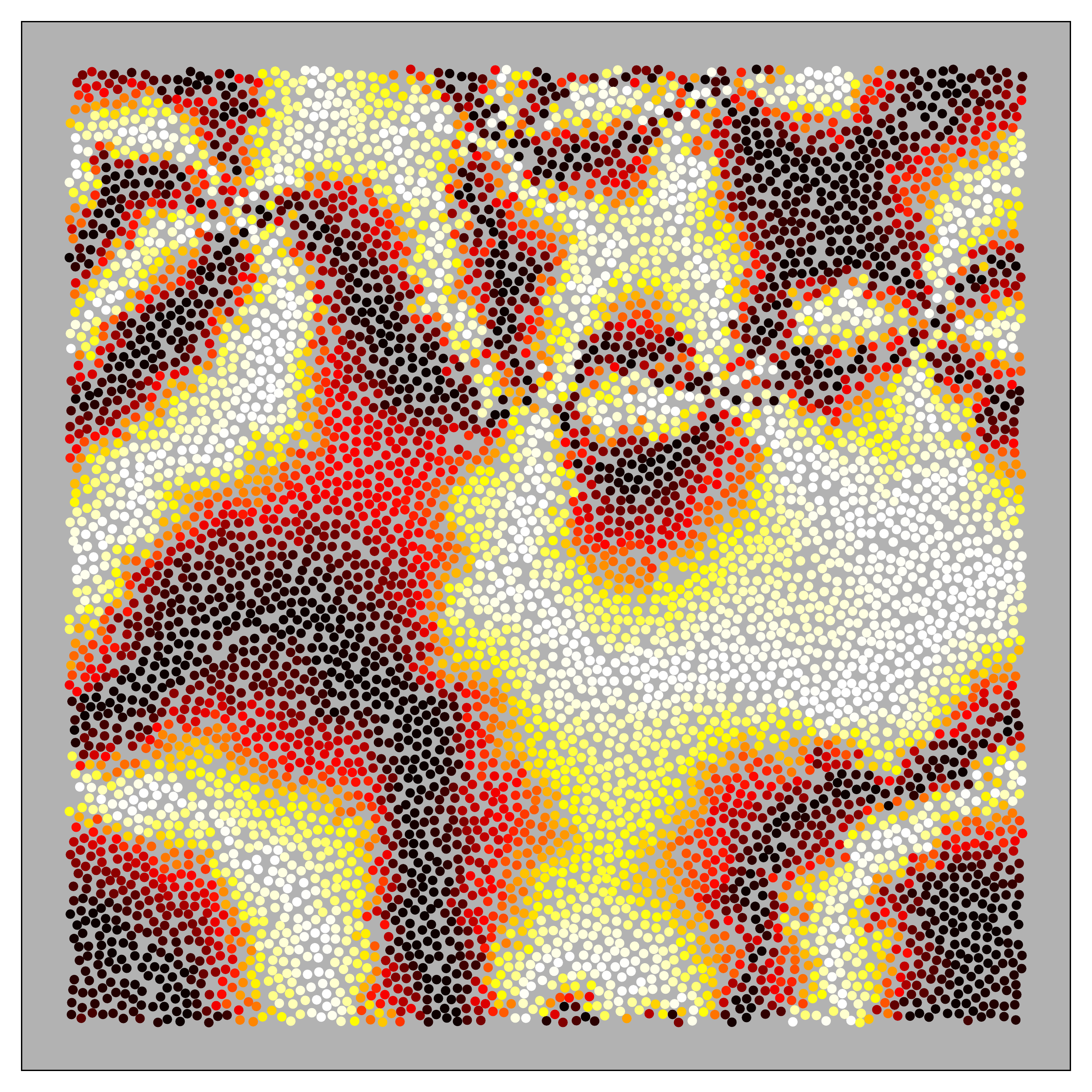}
    } 
    \centering
    \subfloat[]{
    \centering
    \includegraphics[width=0.22\linewidth]{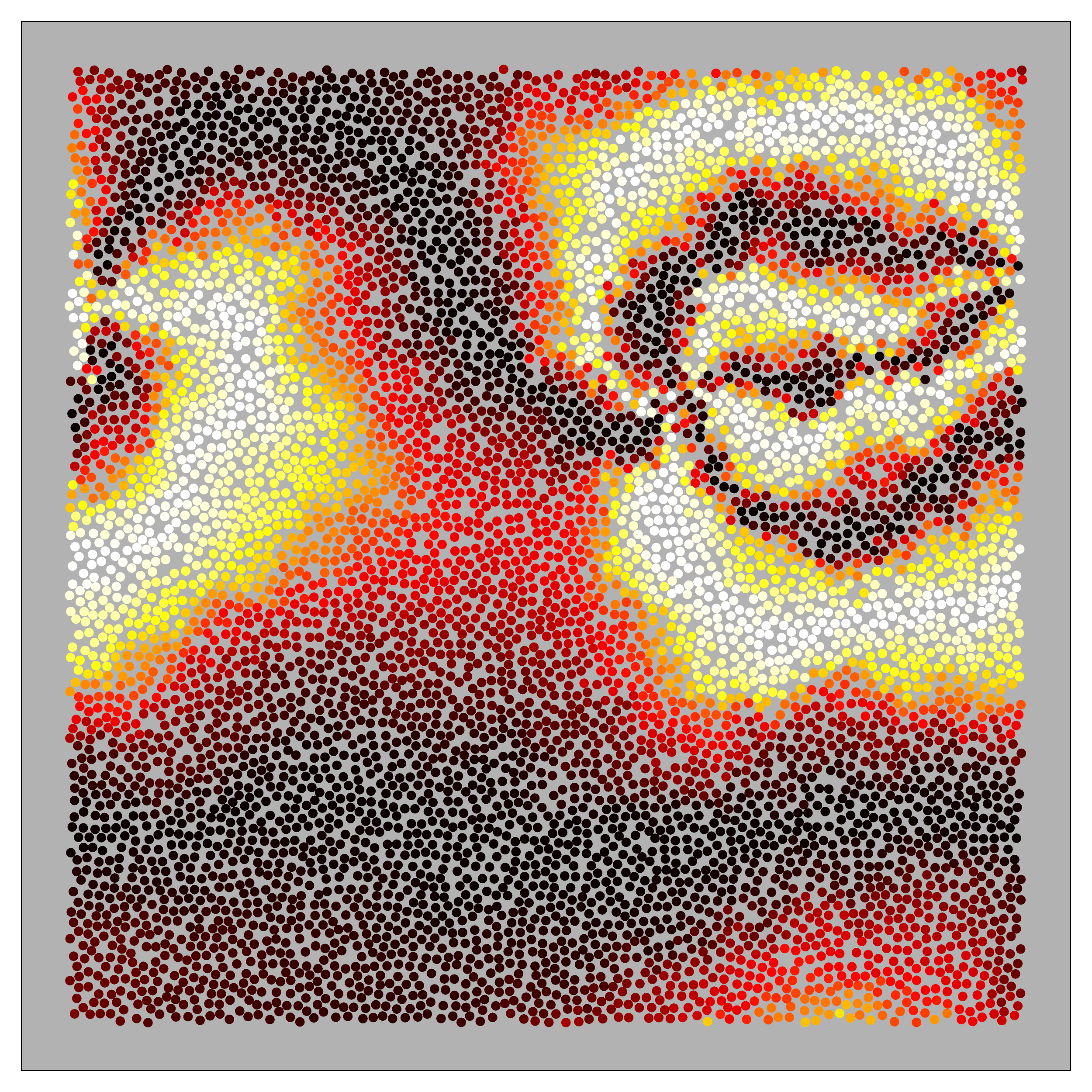}
    }
     \subfloat[]{
    \centering
    \includegraphics[width=0.22\linewidth]{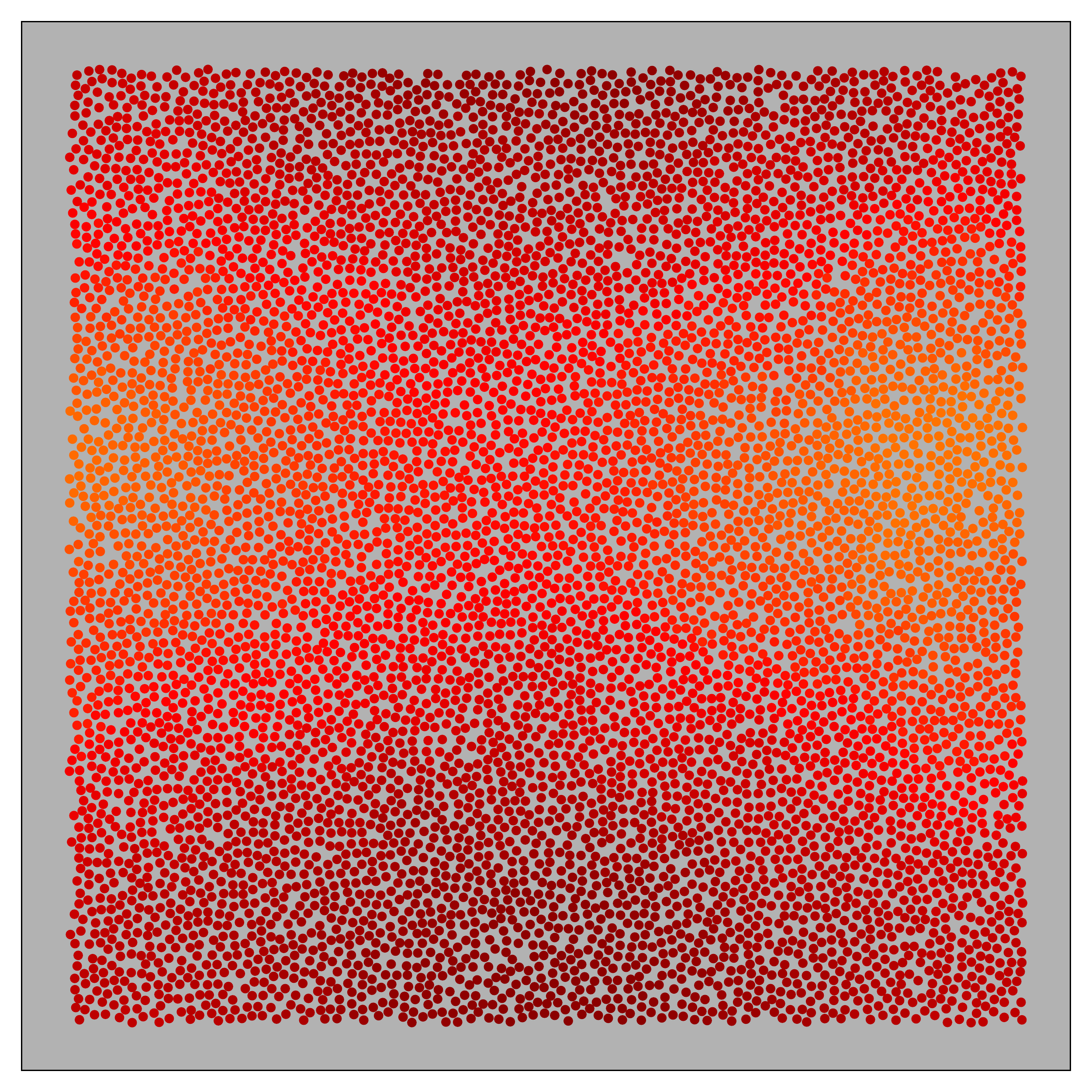}
    } \\
    \centering
    \subfloat[]{
    \centering
    \includegraphics[width=0.22\linewidth]{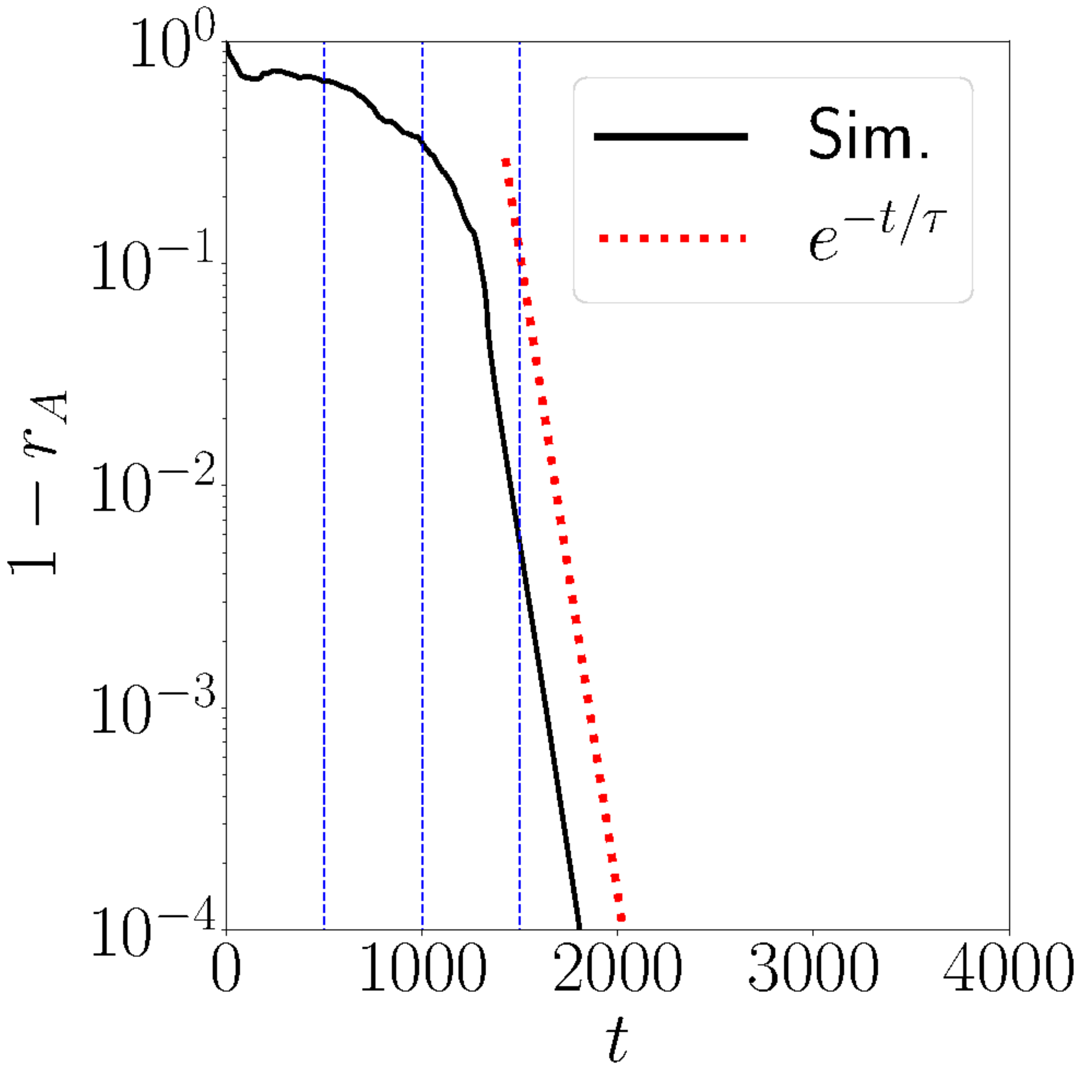}
    }
     \subfloat[]{
    \centering
    \includegraphics[width=0.22\linewidth]{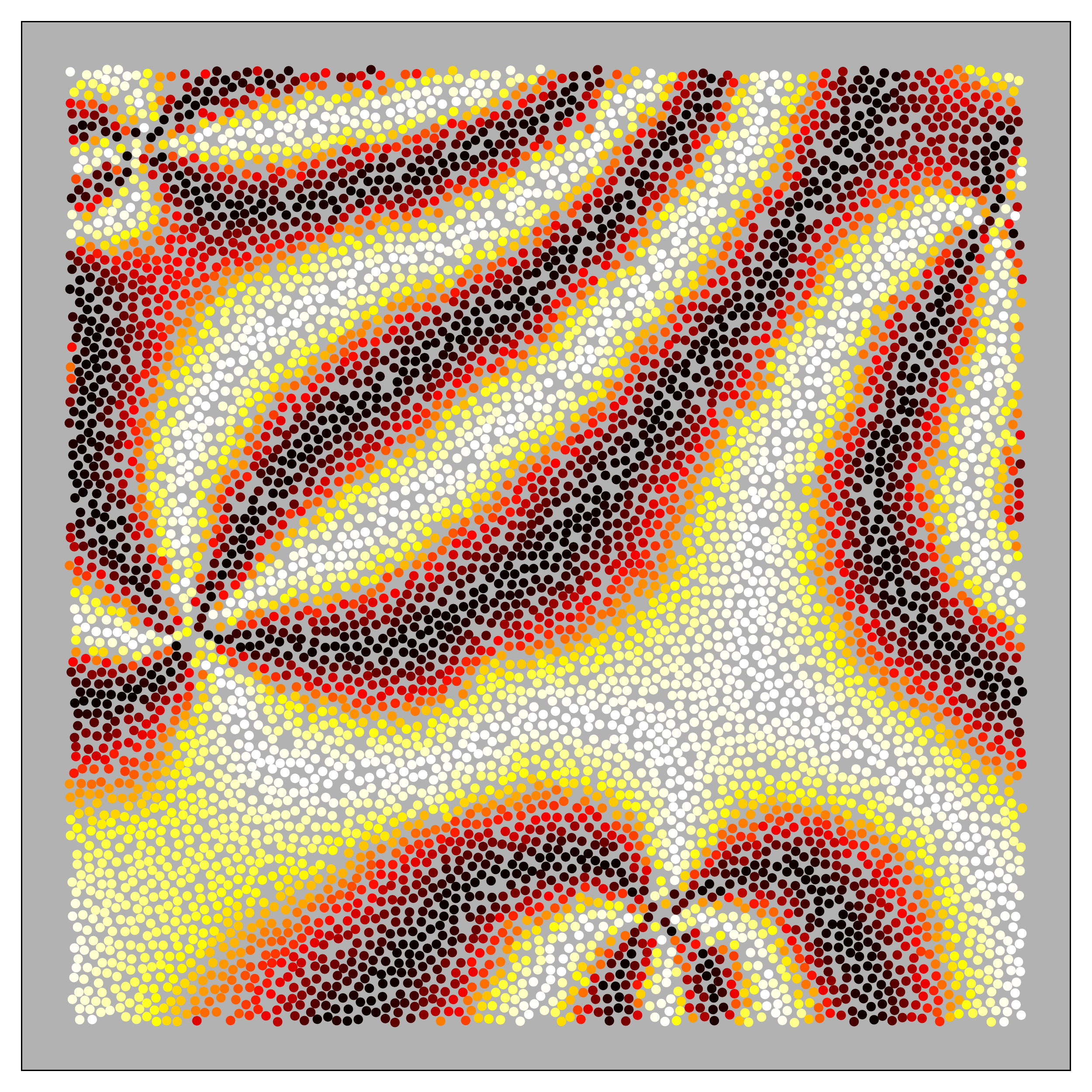}
    } 
    \centering
    \subfloat[]{
    \centering
    \includegraphics[width=0.22\linewidth]{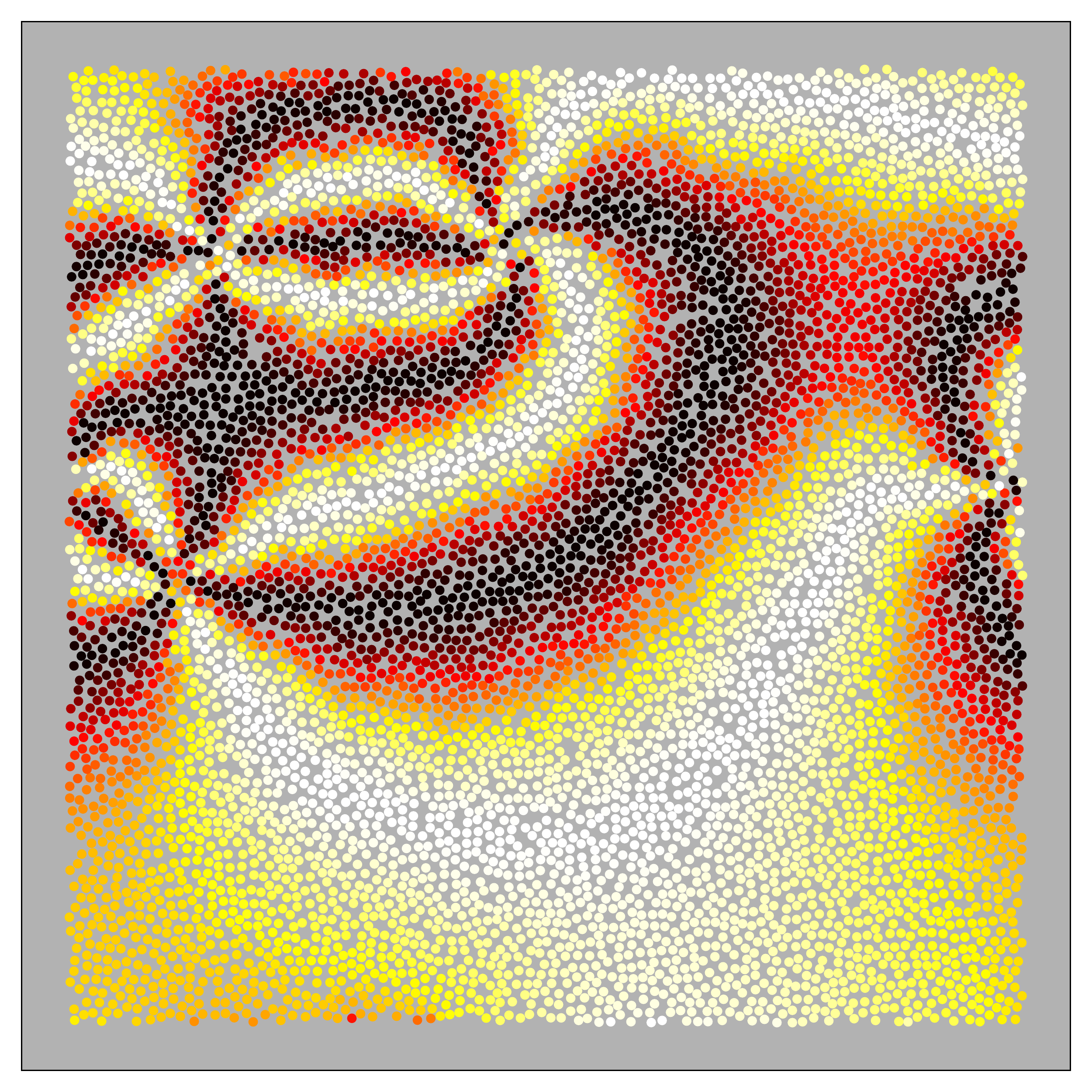}
    }
    \subfloat[]{
    \centering
    \includegraphics[width=0.22\linewidth]{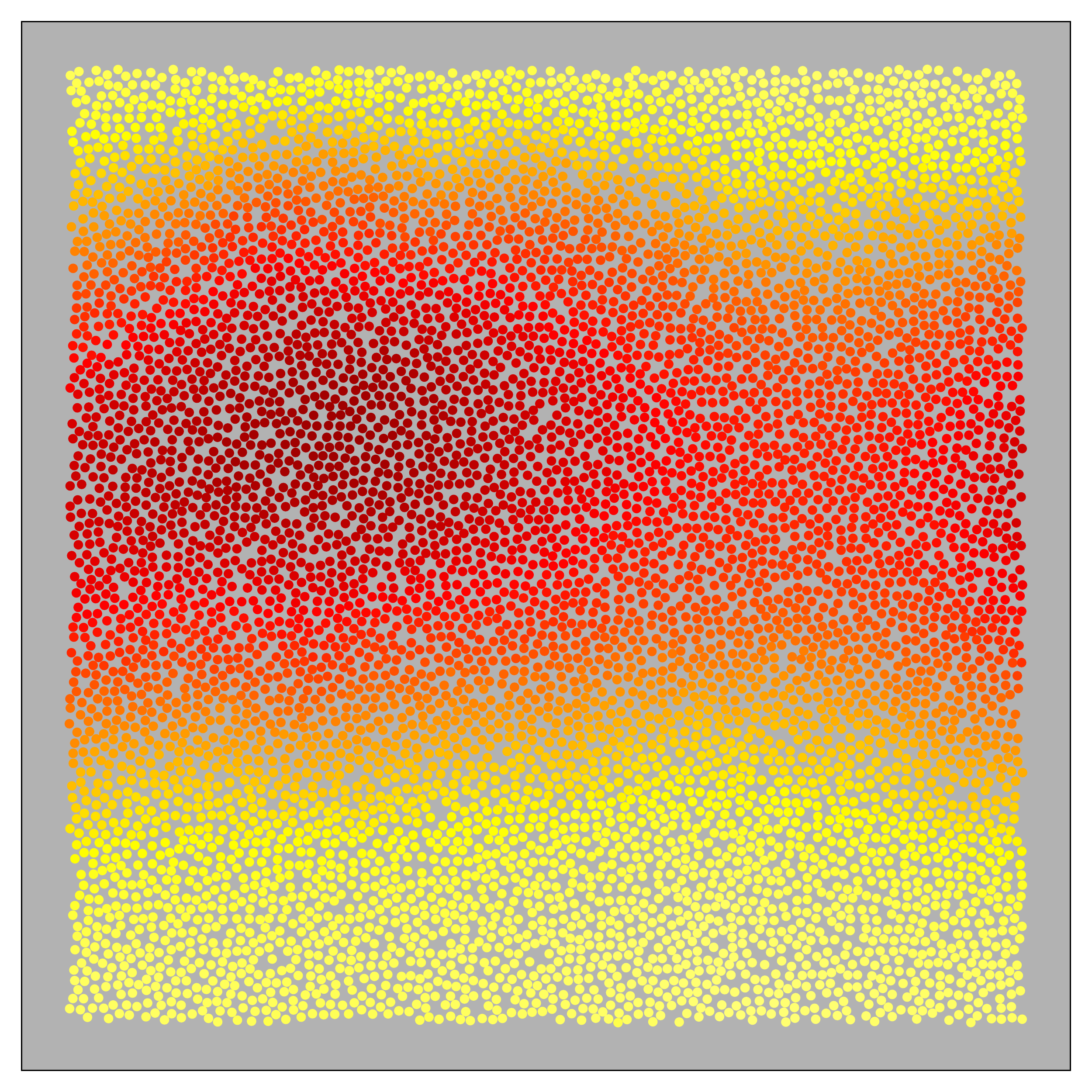}
    }\\
    \centering
    \subfloat[]{
    \centering
    \includegraphics[width=0.22\linewidth]{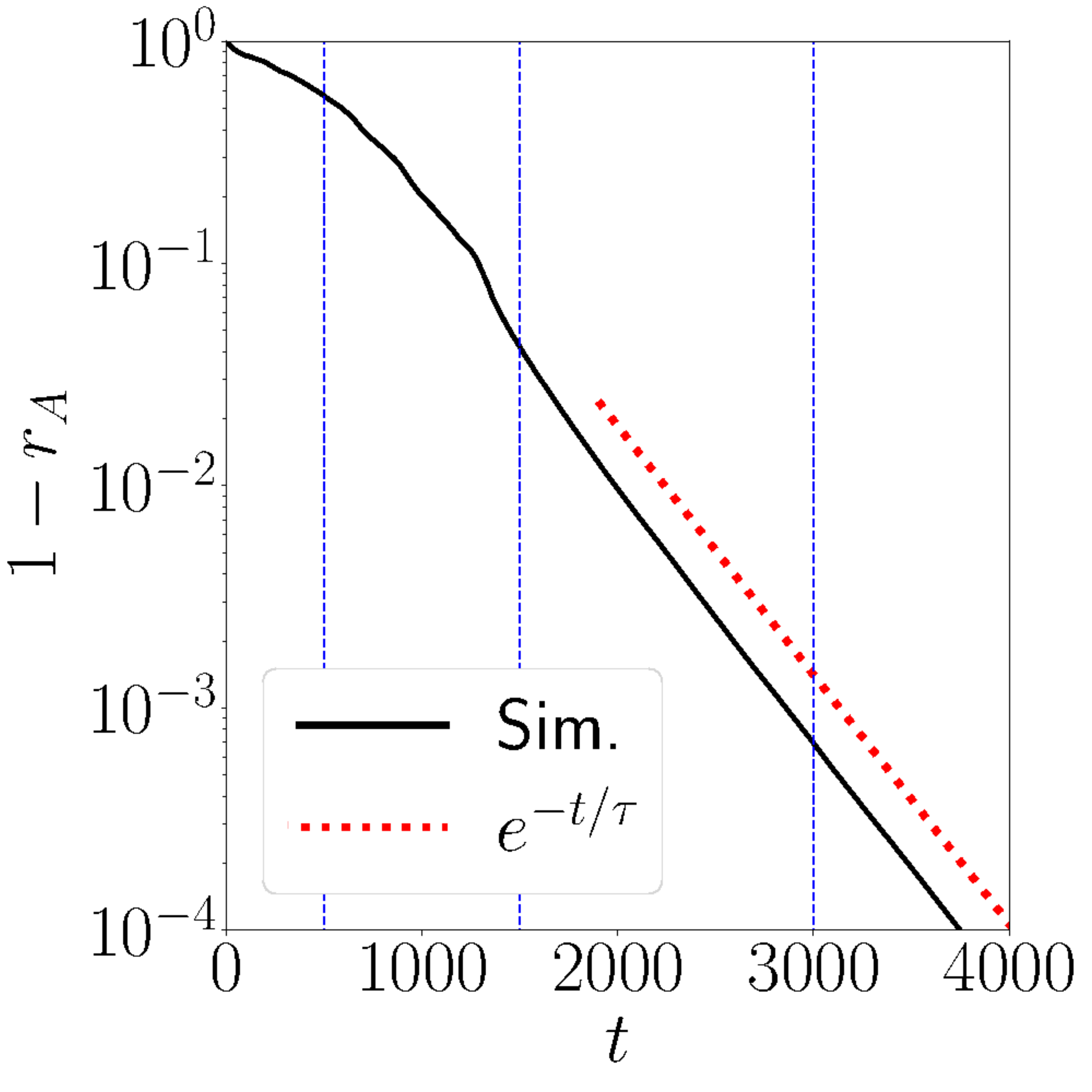}
    }
     \subfloat[]{
    \centering
    \includegraphics[width=0.22\linewidth]{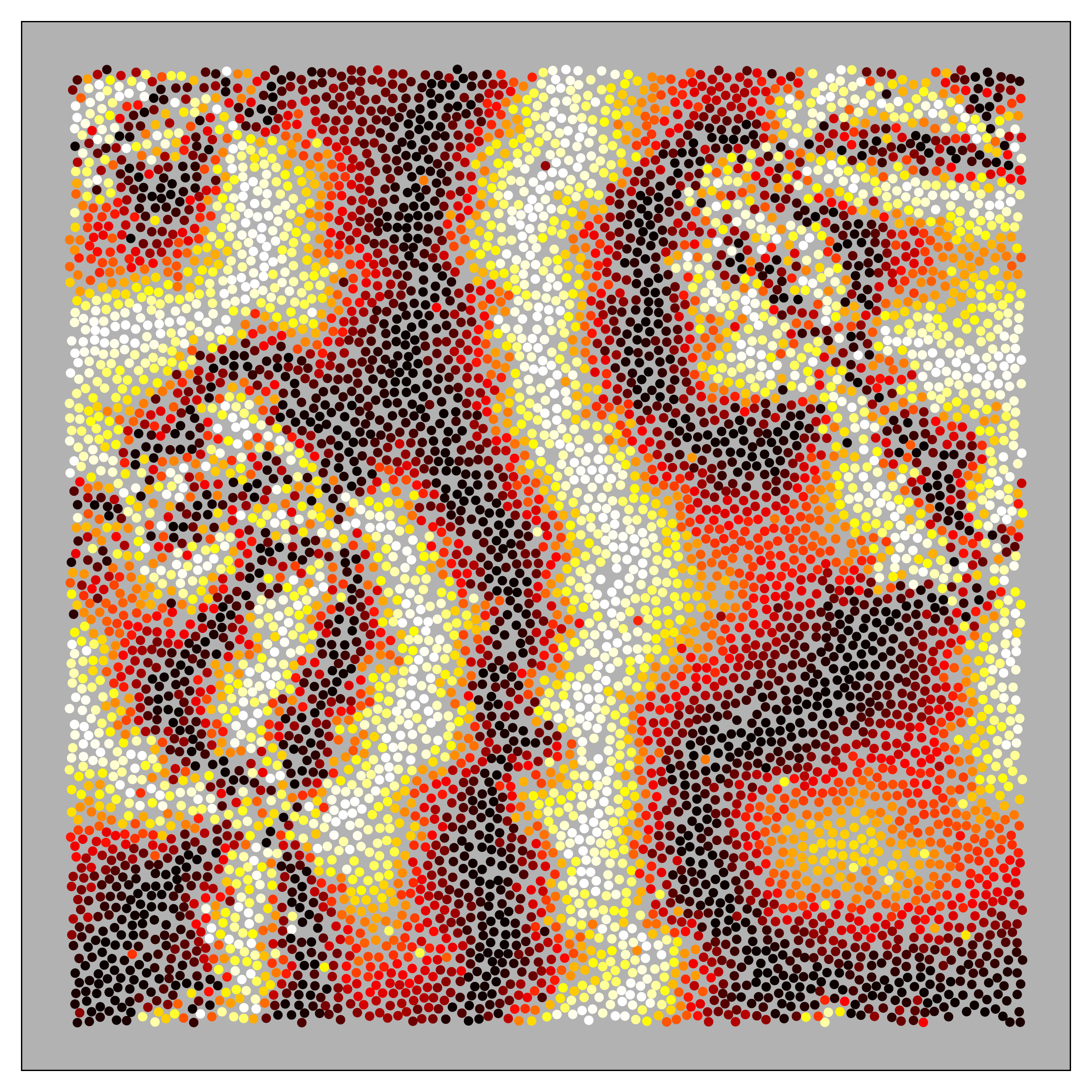}
    } 
    \centering
    \subfloat[]{
    \centering
    \includegraphics[width=0.22\linewidth]{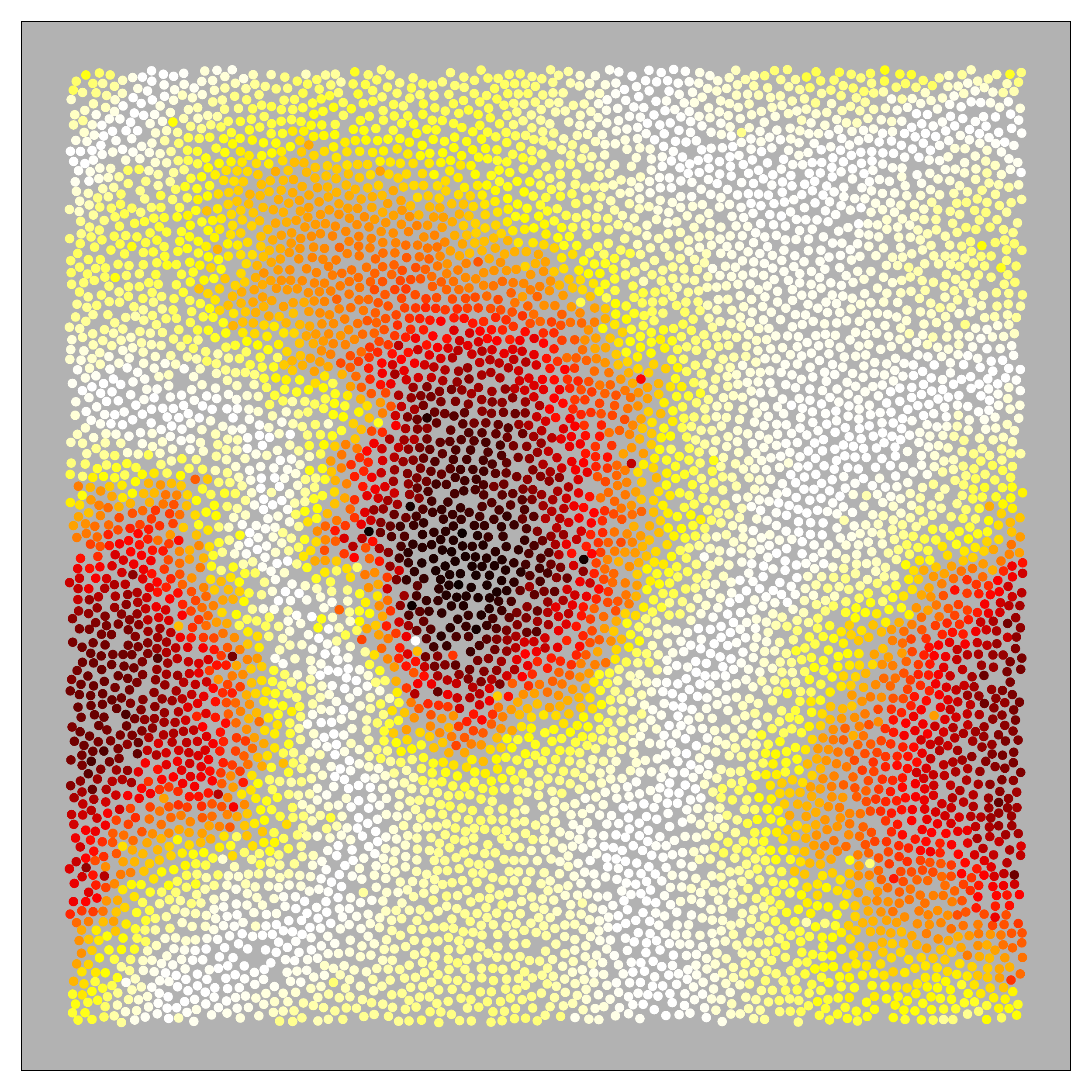}
    }
    \subfloat[]{
    \centering
    \includegraphics[width=0.22\linewidth]{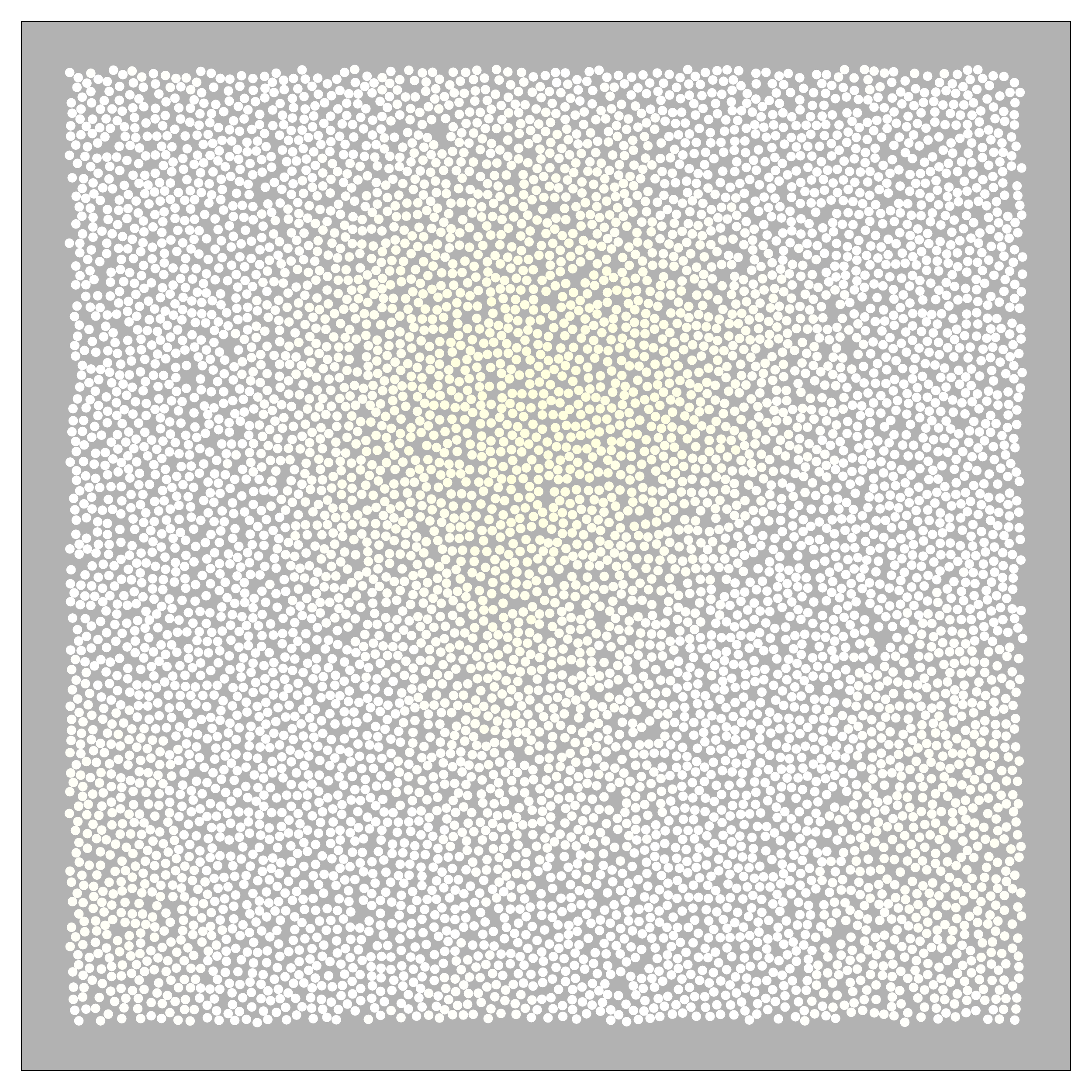}
    }\\
     \subfloat{
    \centering
    \includegraphics[height=0.10\linewidth]{colorbar.pdf}
    }
  
    \vspace{-0.5\baselineskip}
    \caption{Typical instantaneous oscillator configurations obtained in the moderate switching regime at $K = 0.1\K$, which corresponds to $t_r/t_{\theta} = 0.9$. (a)-(d) correspond to $\zeta = 20$ and $\phi=0.05$; (e)-(h) to $\zeta = 20$ and $\phi=0.5$; and (i)-(l) to $\zeta = 0.1$ and $\phi=0.5$. Panels (a),~(e),~(i) depict the late time behavior of $1-r_{\cal A}(t)\sim {\mathrm e}^{-t/\tau}$. The snapshots in each of the rows are taken at times which match those marked by vertical blue lines on panels (a),~(e) and (i), respectively. For all the cases $N=7200$ and $L = 100\sigma$, which corresponds to the packing fraction $\eta\approx 0.57$. Time is given in units of $\tunit$.
     }
    \label{fig:states}
\end{figure*}

Model $\cal I$ exhibits similar mean-field like behavior in the FS regime. Indeed, numerical results obtained for values of coupling constants $0<K,|H|<0.05$ reveal that the time evolution of the order parameter $r_{\cal A}(t)$ has a functional form which agrees semi-quantitatively with the mean-field solution of Ott and Antonsen in Eq.~(\ref{eq:kuramoto_r(t)}), i.e., $r_{\cal A}(t)\sim 1/\sqrt{1+e^{-\kappa (t - t_0)}}$ as shown in figure \ref{fig:fitpanel}(a). The parameter $\kappa$ is related [but not equal] to the inverse of the synchronization time $\tau$, which characterizes the late time exponential behavior of $1-r_{\cal A}(t)$ [see Fig.~\ref{fig:fitpanel}(b)]; $t_0$ defines the location of the inflection point. For the specific case depicted in Fig.~\ref{fig:fitpanel}(a) we find by fitting the numerical data to the mean-field model that $\kappa\approx 1.8\times 10^{-3}$, $t_0\approx 4.1 \times 10^3$.      
Figures \ref{fig:binarysynchronizationpanel}(b)-(d) show snapshots of the system evolution in the FS regime. The network connectivity changes rapidly due to the fast diffusion of the oscillators, and the phases ``experience'' a self-averaged effective all-to-all topology. All phases approach the complete synchronization  with similar rates. No significant fluctuations in the local ordering are visible and the order parameter approximately follows the mean-field dynamics of Eq.(\ref{eq:kuramoto_r(t)}). Similar behavior was found in ref.~\cite{Fujiwara2011} and was named ``global synchronization''.

Different behavior is observed for $0.05<K,|H|$ which is depicted in Figs.~\ref{fig:fitpanel}(c) and (d). Now $t_r/t_{\theta} > 0.45$ such that the local phase dynamics competes with the dynamics of the network reconfiguration. In the limit $t_r/t_{\theta} \gg 1$, named ``slow switching'' regime \cite{levis17} in the language of the dynamical networks, the oscillators are effectively immobile and the dynamics of the phase synchronization can be mapped onto the coarsening relaxation dynamics of the $2D$ $XY$ model after quenching from the infinite temperature to the values below the Kosterlitz-Thouless critical point \cite{levis17,kosterlitz:1974}. The system snapshots in Figs.~\ref{fig:binarysynchronizationpanel}(f)-(h) demonstrate the spatially non-homogeneous structures with locally synchronized regions which grow with time similarly to the coarsening dynamics of $XY$ model. We also observe several topological defects, which appear when growing regions with different average phases meet. Three different time regimes can be clearly identify in $r_{\cal A}(t)$ in Fig.~\ref{fig:fitpanel}(c). Approximately linear regime for $t\leq200$ where all the curves approximately collapse on top of each other. This is the regime when the locally coherent regions appear in different locations and grow in an overall disordered background. When these regions start touching, the orientational topological defects emerge and initiate the second dynamical regime for $200\leq t \leq 1500$. This regime is dominated by the motion and annihilation of the topological defects with opposite winding numbers, which in turn gives rise to strongly fluctuating $r_{\cal A}(t)$ curves. At later times, $t\geq 1500$, final homogenisation of the phases takes place, and during this process $r_{\cal A}(t)$ can be very well approximated by the mean-field profile in Eq.~(\ref{eq:kuramoto_r(t)}), see red dashed curves in the inset of Fig.~\ref{fig:fitpanel}(c).

\begin{figure}[h!]
    \centering
    \captionsetup{position=top,justification=raggedright,singlelinecheck=false}
    \subfloat[]{
    \centering
    \includegraphics[width=0.48\linewidth]{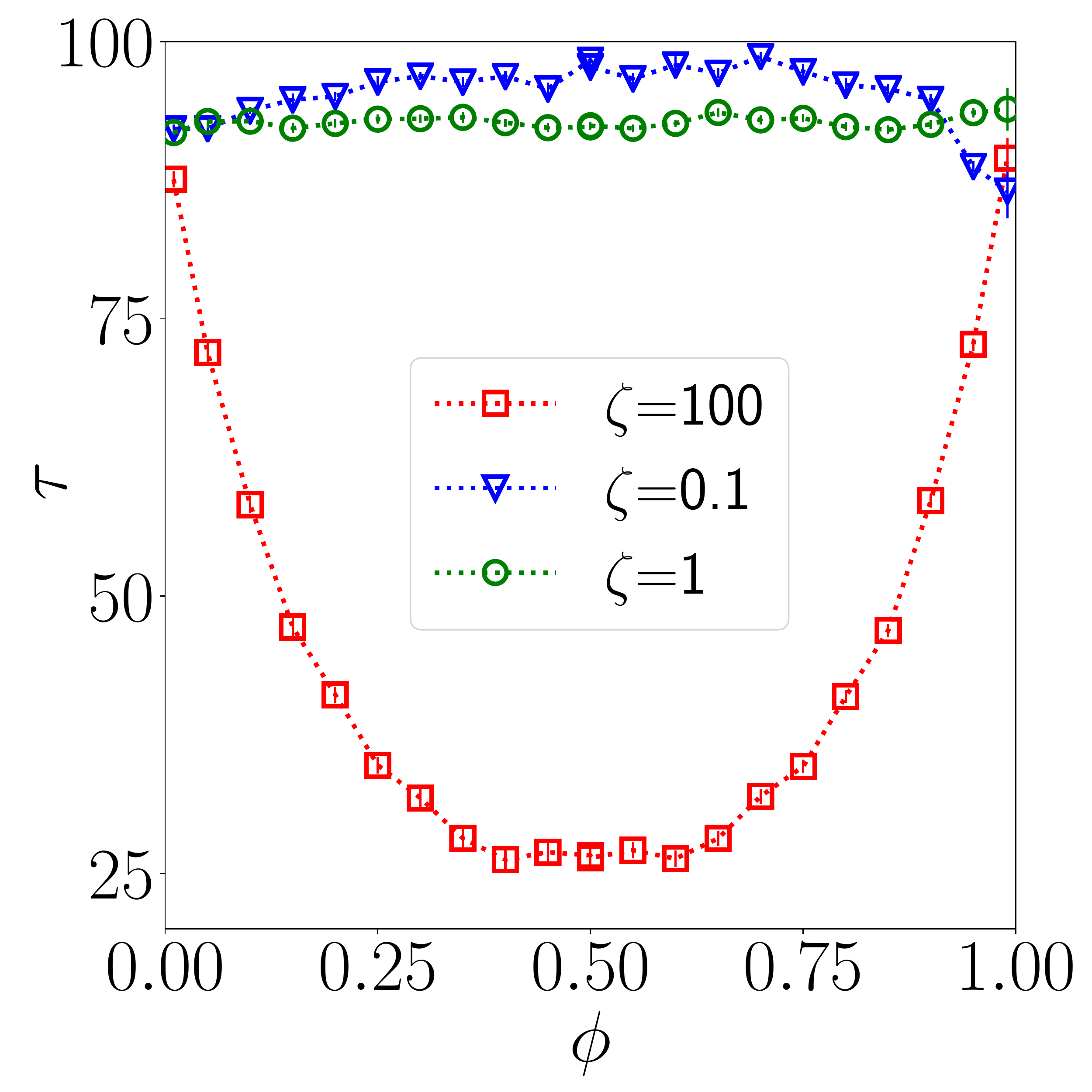}
    }
     \subfloat[]{
    \centering
    \includegraphics[width=0.48\linewidth]{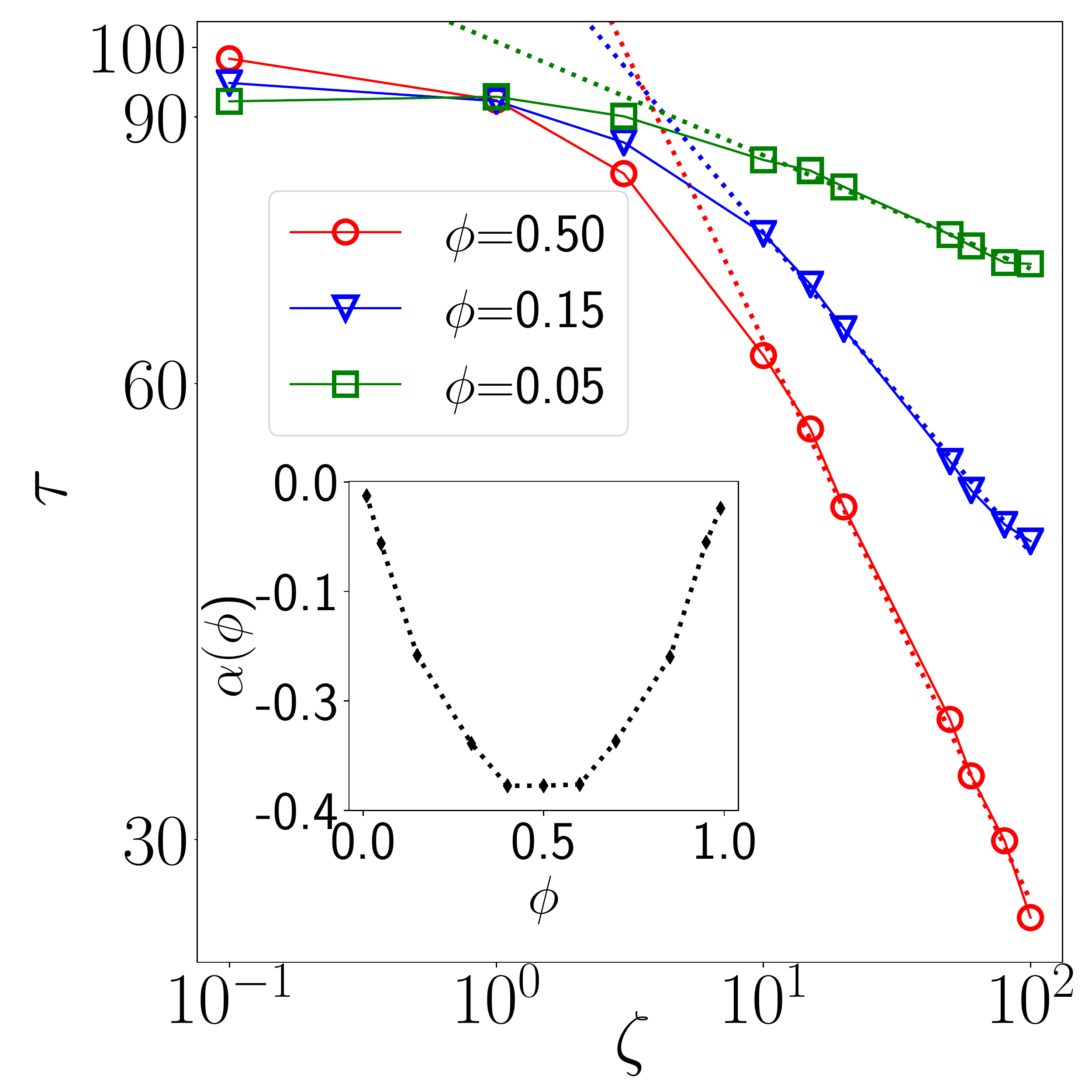}
    }
    \captionsetup{justification=centerlast}
    \vspace{-0.5\baselineskip}
    \caption{(a) Synchronization time $\tau$ as a function of $\phi$ for several values of $\zeta$. (b) 
    $\tau$ as a function of $\zeta$ for several values of $\phi$. Solid dotted lines represent fits to a power-law $\tau \propto \zeta^{\alpha(\phi)}$, only data points for $\zeta >10$ are fitted. The resulting exponent $\alpha$ is shown as a function of $\phi$ in the inset. In all the cases $N = 1800$, $L=50\sigma$, $K=0.1\K$. 
    The results are obtained after averaging over 100 independent runs.}
    \label{fig:tauf}
\end{figure}

As we discussed above, in the FS regime the system exhibits a mean-field behavior, and it is expected that for pure ${\cal A}-$system, $\phi=0$, $\kappa \propto K$ and $t_0 \propto K^{-1}$ \cite{Fujiwara2011}. The addition of the oscillators of type ${\cal B}$ changes this behavior such that  both $\kappa$ and $t_0$ become non-trivial functions of $K$, $\zeta$, $\phi$ and $\eta$. We start by plotting in Fig.~\ref{fig:KappaT0phi} $\kappa$ and $t_0$ as functions of $\phi$ at several values of $\zeta$ and at constant $K$ and $\eta$. All the systems represented belong to the FS regime. Surprisingly, $\kappa$ and $t_0$ do not depend of the number of ${\cal B}-$oscillators introduced [within the numerical uncertainty] for $\zeta = 1$. In this case both subpopulations synchronize at the same rate, but with the average phases that are diametrically opposed to each other. By increasing $\zeta$ the phase repulsion between locally synchronized  ${\cal A}-$ and ${\cal B}-$domains increases. This evokes a kind of a positive feedback mechanism when a given locally coherent cluster acquires new members mostly because those are being repelled by a neighbouring coherent cluster of another type. This effect is more pronounced the larger the two oppositely synchronized clusters are, such that eventually the global synchronization is driven by the repulsion, rapidly [for large $\zeta$] driving the two subpopulations into diametrically opposed states, decreasing the effective synchronisation time $\tau\sim\kappa^{-1}$ of ${\cal A}-$oscillators, see blue triangles and red squares in Fig.~\ref{fig:KappaT0phi}(a). The location of the inflection point $t_0$ also decreases with increasing $\zeta$. The results show that there is an optimum composition of the mixture $\phi_{min} \approx 0.6$ at which $\kappa^{-1}$ and $t_0$ attain their minimum values. Based on the symmetry arguments, one would expect rather symmetric shapes for both $\tau (\phi)$ and $t_0(\phi)$ curves, with the respective extrema located at $\phi_{min} \approx 0.5$, similarly to the case of the slow switching regime discussed below [see Fig.~\ref{fig:tauf}]. We relate this slight asymmetry in the present case to the finite size effects. 

\begin{figure}[h!]
    \centering
    \captionsetup{position=top,justification=raggedright,singlelinecheck=false}
    \subfloat[]{
    \centering
    \includegraphics[width=0.48\linewidth]{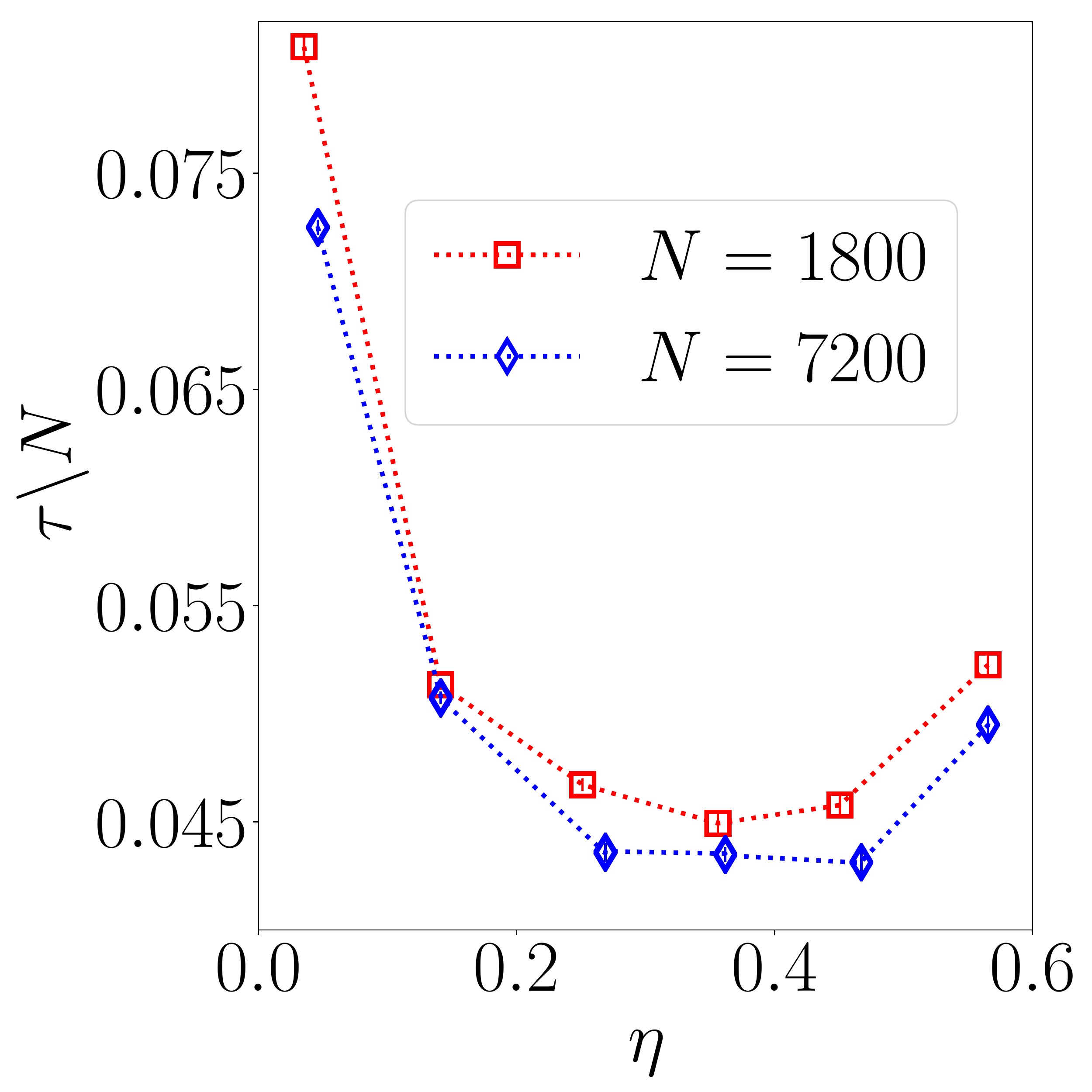}
    }
     \subfloat[]{
    \centering
    \includegraphics[width=0.48\linewidth]{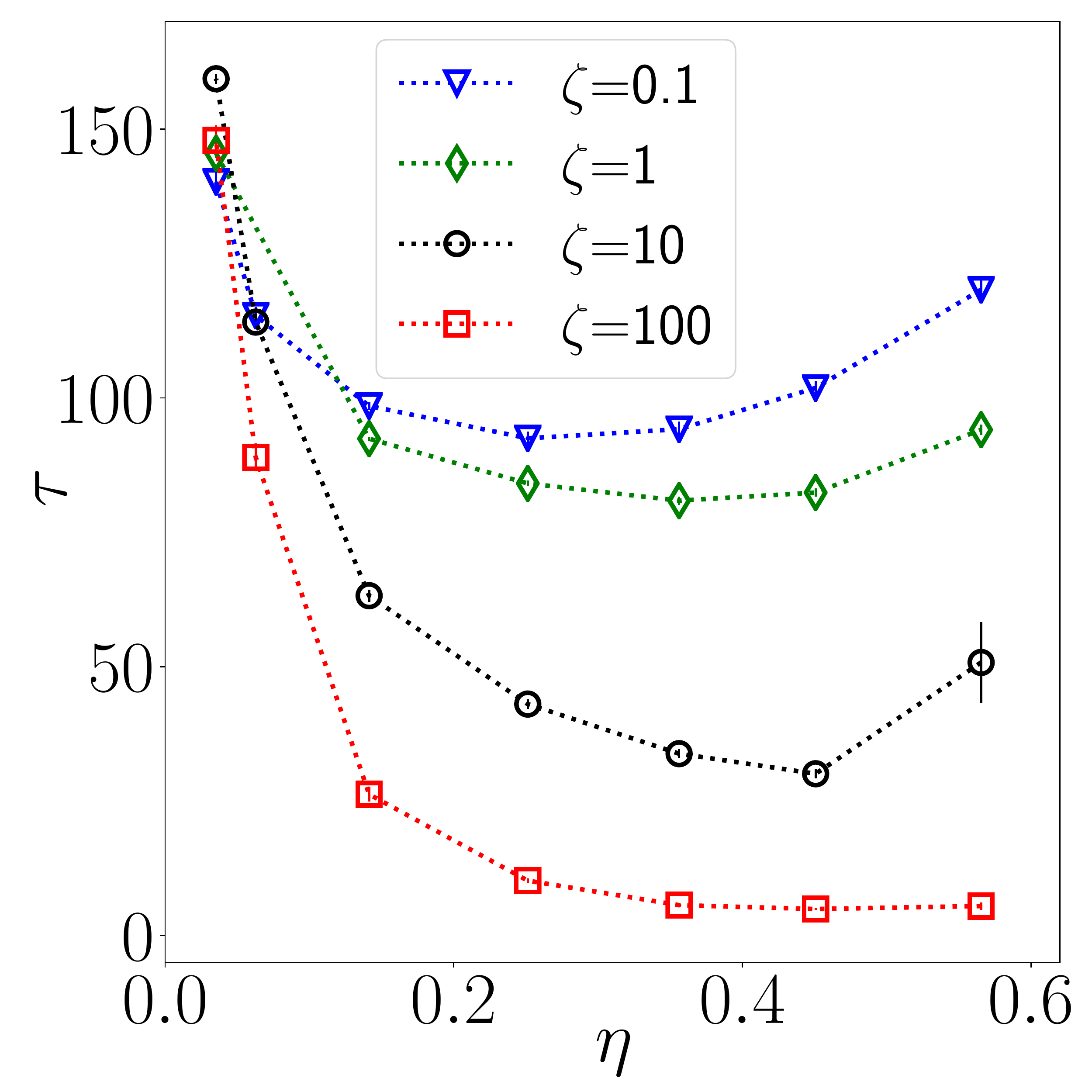}
    } 
    \captionsetup{singlelinecheck=true,justification=centerlast}
    \vspace{-0.5\baselineskip}
    \caption{(a) Synchronization time $\tau$ as a function of the packing fraction $\eta$ for two values of $N$. Rescaling the vertical axis by $N^{-1}$ results in an approximate data collapse. The packing fraction is varied by changing the system size $L$; $K=0.1\K$, $\zeta=1$, $\phi=0.5$. (b) $\tau$ as a function of $\eta$ for several values of $\zeta$. $K=0.1\K$, $\phi=0.5$, $N=1800$.
    $\tau$ is given in units of $\tunit$. The results are obtained after averaging over 100 independent runs.
     }
    \label{fig:escalar}
\end{figure}

Figure \ref{fig:KappaT0H} summarizes the dependence of $\kappa$ and $t_0$ on the interactions asymmetry $\zeta$ for several values of $\phi$. We emphasize that the presence of just a 5\% of controlling agents of type ${\cal B}$ with the interaction asymmetry $\zeta = 10$ reduces $\kappa^{-1}$ of the control population ${\cal A}$ by a factor of 2, see green circles in Fig.~\ref{fig:KappaT0H}(a). 
% shows an increase of the value of $\kappa$ as the value of $\zeta$ increases and that there is a dependence on  $\phi$. We see the opposite happen when when plotting $t_0$ as a function of $zeta$ for different values of $\phi$, see figure \ref{fig:KappaT0H}(b).

In the moderate switching regime, when $K$ or $|H|$ is larger than $0.05\K$ we compute the synchronization time $\tau$ by fitting the late time behaviour of $1-r_{\cal A}(t)$ to the exponential time decay $\propto {\mathrm e}^{-t/\tau}$, as is demonstrated in Fig.~\ref{fig:fitpanel}(d) and in the first column of Fig.\ref{fig:states}. In Fig.~\ref{fig:states} we compare the synchronization dynamics of a ${\cal A}-$rich system at $\phi=0.05, \zeta=20$ [1st row], and $50/50$ mixtures of ${\cal A}-$ and ${\cal B}-$type oscillators at $\zeta = 20$ [2nd row] and $\zeta = 0.1$ [3rd row].
The results demonstrate that by adding the repulsive control oscillators we can either decrease [$\zeta = 20$] or increase  [$\zeta = 0.1$] the synchronization time $\tau$ of the target ${\cal A}-$subpopulation. Similar to the case depicted in Fig.~\ref{fig:binarysynchronizationpanel}, here we also observe the coarsening dynamics driven by vortex-antivortex annihilation reminiscent of the $2D$ $XY$ model. The defects are more pronounced in these simulations because we have used a larger number of oscillators.

\begin{figure}[h!]
    \centering
    \captionsetup{position=top,justification=raggedright,singlelinecheck=false}
    \subfloat[]{
    \centering
    \includegraphics[width=0.48\linewidth]{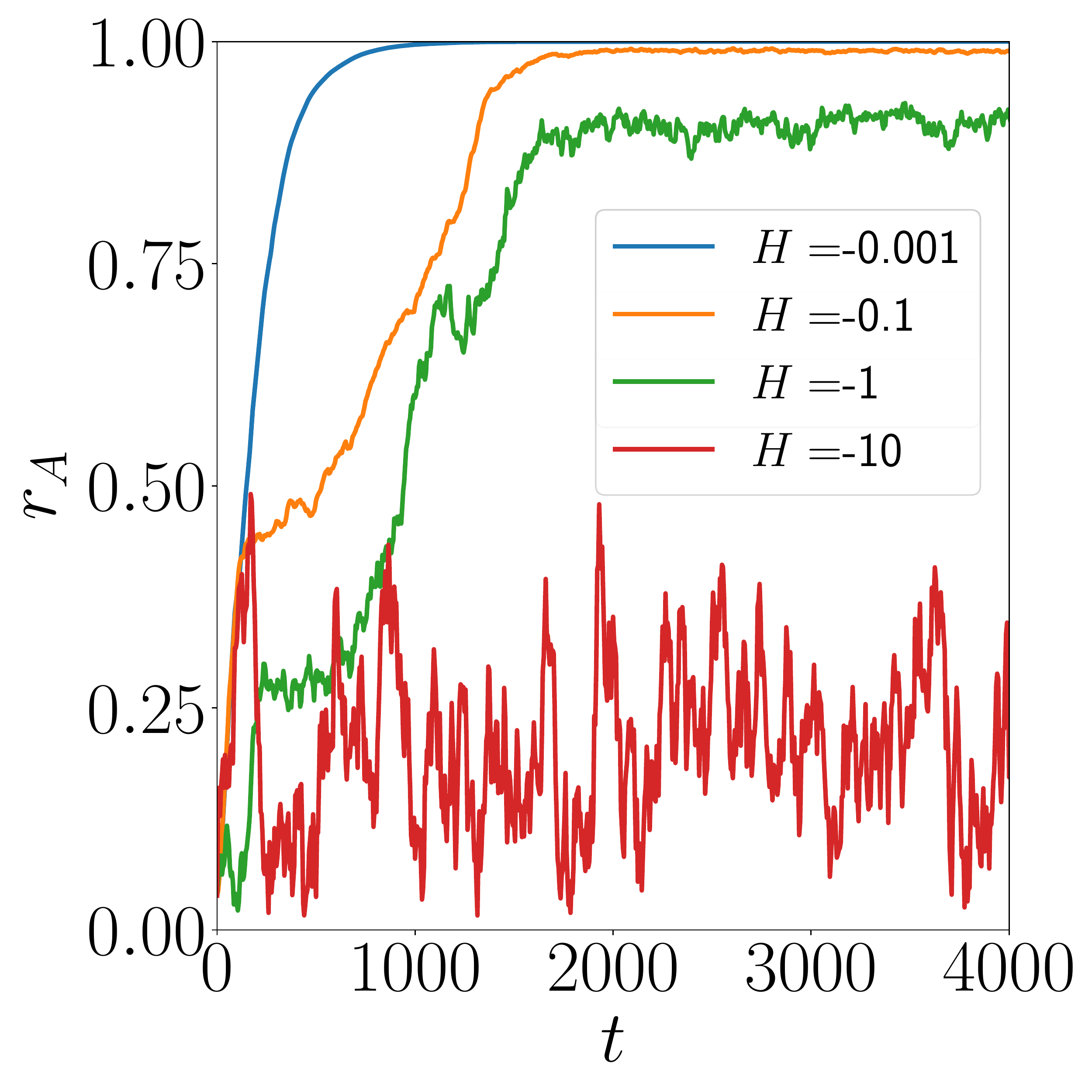}
    }
     \subfloat[]{
    \centering
    \includegraphics[width=0.48\linewidth]{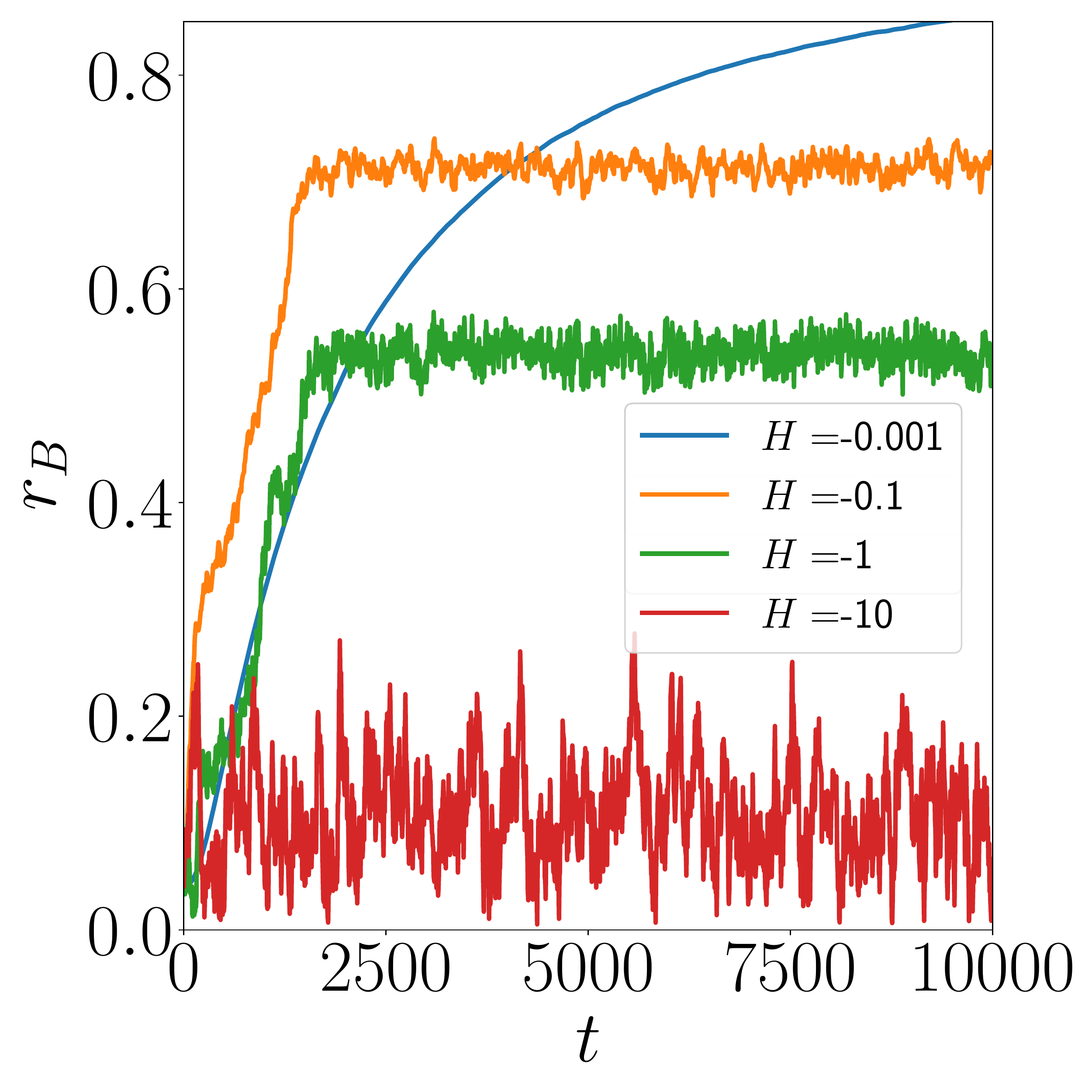}
    } 
    \vspace{-0.5\baselineskip}
    \captionsetup{justification=centerlast}
    \caption{(a)  $r_{\cal A}$, (b) $r_{\cal B}$ as functions of time for several values of $H$. $N=1800$, $L=100\sigma$, $K=0.001 \K$, and $\phi=0.5$. ${\cal A}-$oscillators exhibit complete synchronization with $r_{\cal A}^*=1$, and ${\cal B}-$oscillators we find $r_{\cal B}^*>0.7$ for 
     $|H| < 0.1\K$.  $r_{\cal A}^*$  decreases systematically with $|H|$ for $|H|>0.1\K$. Time is shown in units of $\tunit$. }
    \label{fig:modelIIsynchronization}
\end{figure}

We find that in moderate switching regime the synchronization time $\tau$, in contrast to the case of the FS regime shown in Fig.~\ref{fig:KappaT0phi}(a), is a symmetric function of $\phi$ with the minimum [maximum] at $\phi=0.5$ for $\zeta>1$ [$\zeta<1$], as shown in Fig.~\ref{fig:tauf}(a). This behavior is related to the underlying symmetry of the model: the swapping of the identities of the oscillators combined with the replacement of $\phi$ with $1-\phi$ does not affect the model behavior. We emphasize, however, that the mechanisms driving the synchronization of ${\cal A}-$oscillators are different for small and large $\phi$. Indeed, for $\phi\ll 1$ the synchronization of ${\cal A}-$subpopulation is driven by the intrapopulation attractions, while for $\phi\approx 1$ it is mostly the repulsion from the coherent ${\cal B}-$subpopulation which promotes the synchronization of ${\cal A}-$oscillators. For example, for the system of Fig.~\ref{fig:tauf}(a) the average distance between ${\cal A}-$oscillators is $\approx 5.3\sigma$ at $\phi = 0.95$ which is almost twice the interaction range, therefore the contribution of the attractive intrapopulation interactions is insignificant. 

\begin{figure}[h!]
    \centering
    \captionsetup{position=top,justification=raggedright,singlelinecheck=false}
    \subfloat[]{
    \centering
    \includegraphics[width=1\linewidth]{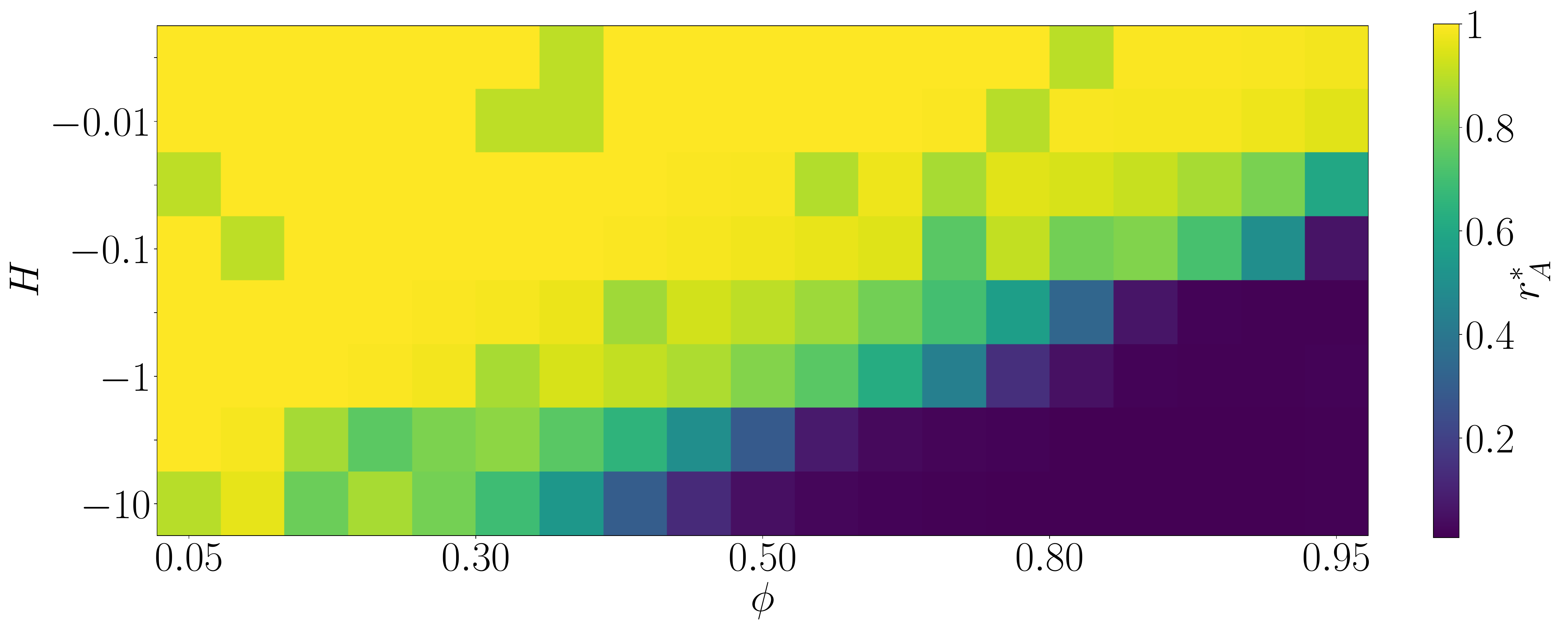}
    }\\
     \subfloat[]{
    \centering
    \includegraphics[width=1\linewidth]{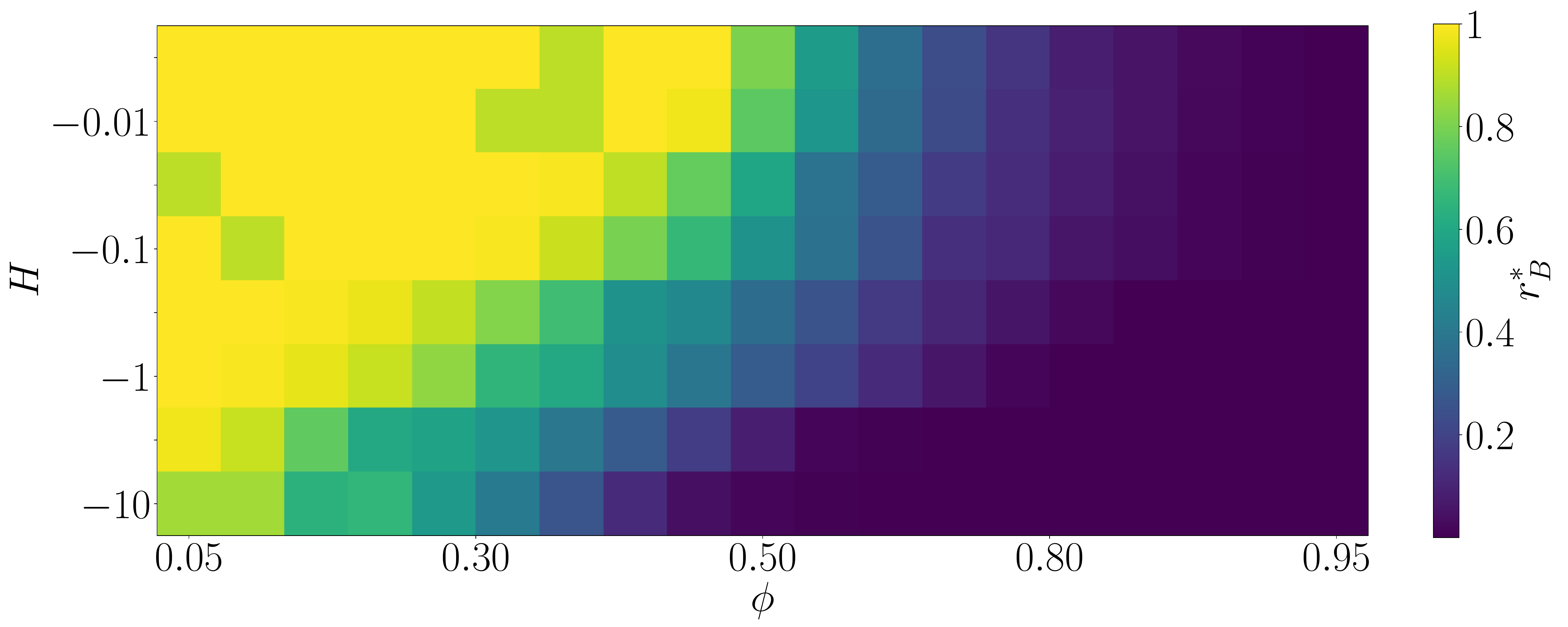}
    } 
    \captionsetup{justification=centerlast}
    \vspace{-0.5\baselineskip}
    \caption{Heat map of the late time values of the order parameter for (a) ${\cal A}-$oscillators and (b) ${\cal B}-$oscillators. The maps show the existence of configurational transitions from fully synchronized $r_{\sigma}^*=1$ to disordered $r_{\sigma}^*\ll 1$ configurations driven either by $\phi$ or $H$. All simulations were carried out at $K=0.1\K$, $N=1800$, $L=100$.}
    \label{fig:diagramaA}
\end{figure}

We observe that $\tau$ does not depend [up to numerical uncertainty] on the composition of the mixture at $\zeta=1$. Our numerical results also suggest that for $\zeta\gtrsim 10$ the synchronization time as a function of $\zeta$ has a power-law form with an exponent $\alpha$ that depend on $\phi$ and with the minimum at $\phi=0.5$, as shown in Fig.~\ref{fig:tauf}(b). 

In the slow switching regime, when the network is effectively static, one can argue using the dynamic scaling hypothesis \cite{hohenerg:1977} that the synchronization time $\tau\sim N$ \cite{levis17}. Surprisingly, we find that this scaling also holds approximately in the moderate switching regime. In Fig.~\ref{fig:escalar}(a) we plot $\tau/N$ as a function of the packing fraction for two different $N$, showing an approximate data collapse onto a single master curve. Figure~\ref{fig:escalar}(b) also shows that there is a $\zeta-$dependent optimum value $\eta_{min}(\zeta)$ of the packing fraction at which $\tau$ is minimal. At the lowest value of the packing fraction considered, $\eta=0.06$, the average distance between the oscillators is $\approx 3.6$ which is above the range of the interactions between the phases, leading to the increase of $\tau$. Contrary, at higher values of $\eta$ the mobility of the oscillators decreases due to the steric jamming of the system, resulting in a slower mixing and larger $\tau$. Indeed, we find numerically that for this system the diffusion coefficient varies over an order of magnitude for $0.06 \leq \eta \leq 0.6$. Surprisingly, the effect of the steric jamming can be overcome by increasing $\zeta$, see red squares in Fig.~\ref{fig:escalar}(a) 

\begin{figure}[h!]
    \captionsetup{position=top,justification=raggedright,singlelinecheck=false}
    \centering
    \subfloat[]{
    \centering
    \includegraphics[width=0.48\linewidth]{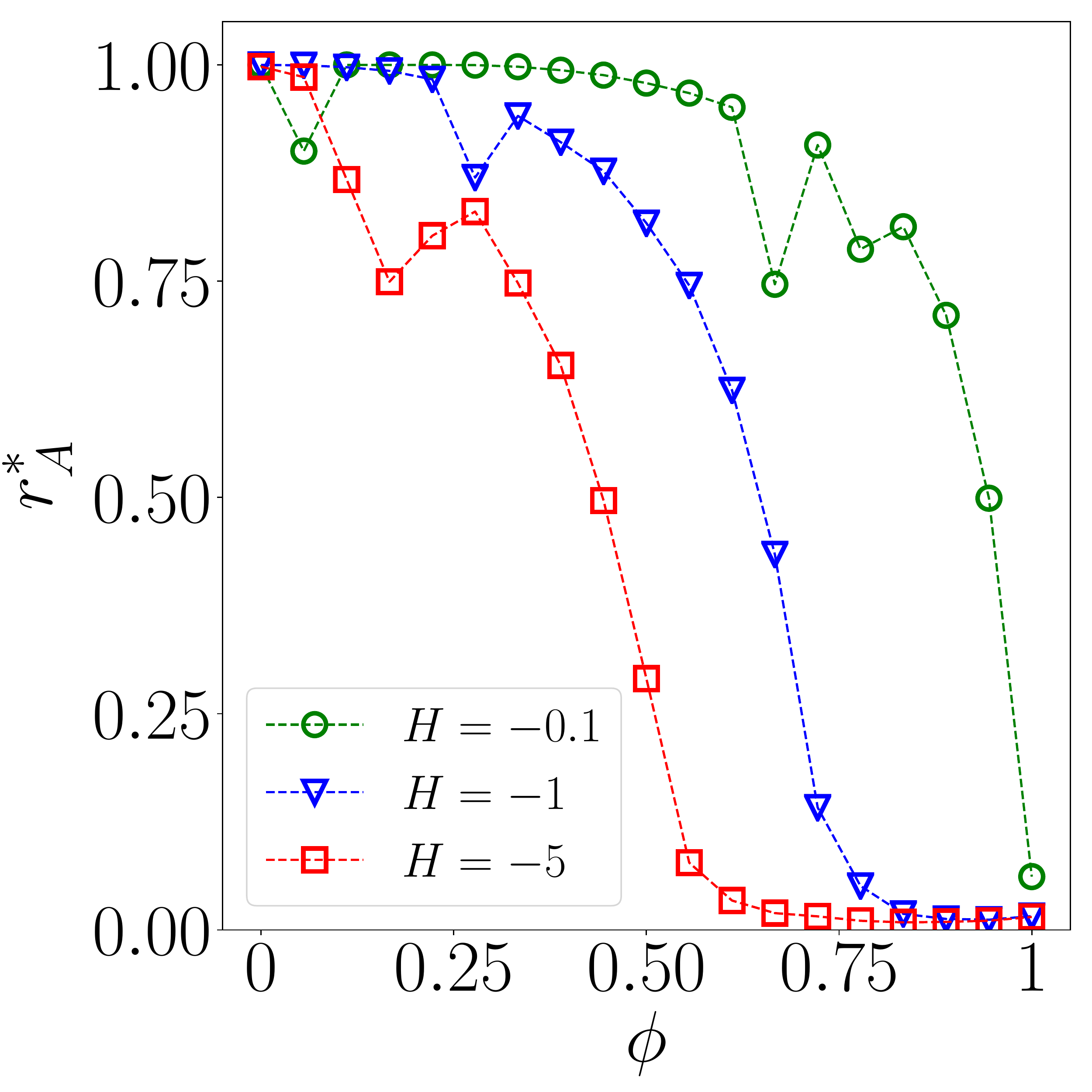}
    }
     \subfloat[]{
    \centering
    \includegraphics[width=0.48\linewidth]{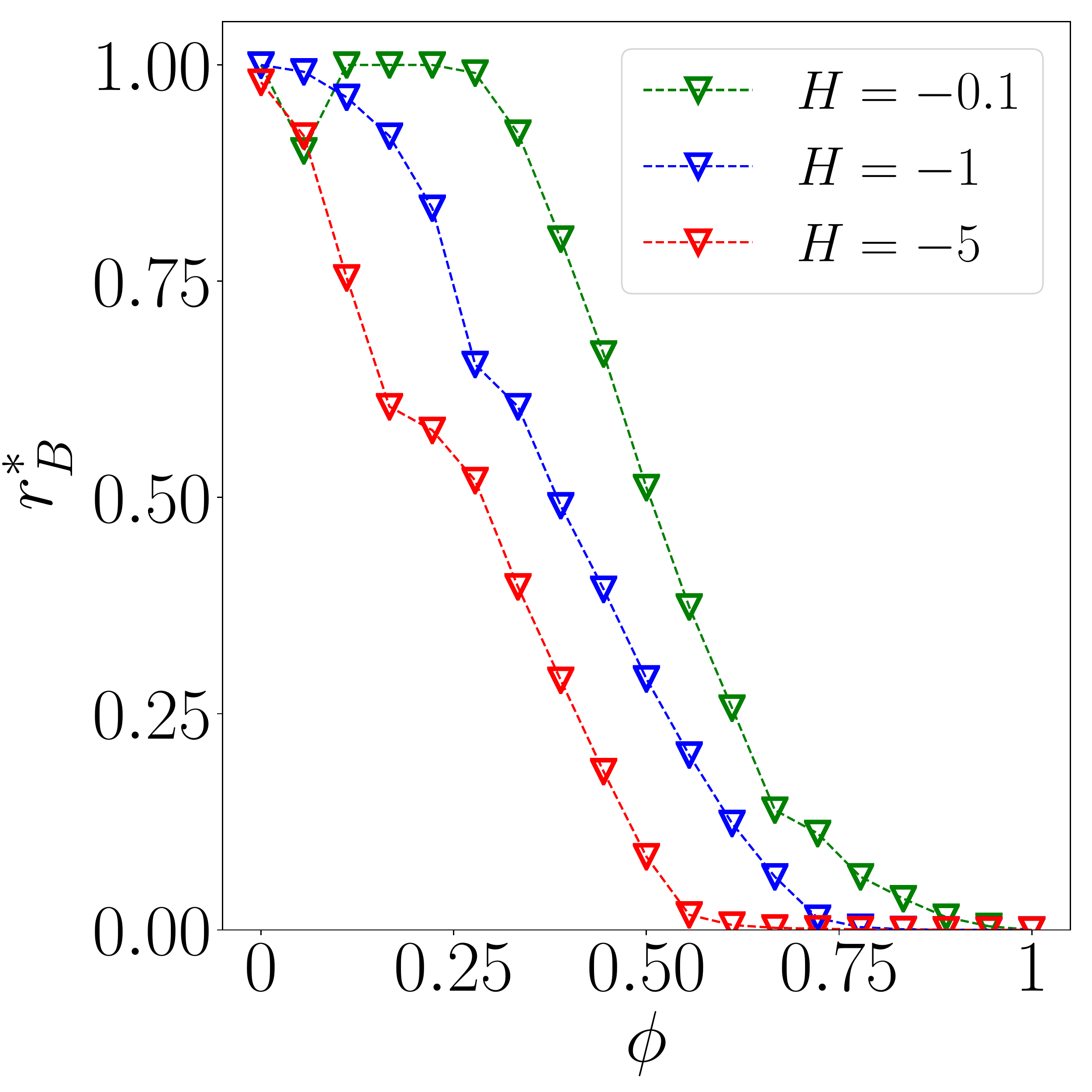}
    } 
    \vspace{-0.5\baselineskip}
    %\captionsetup{justification=centerlast}
    \caption{(a) $r_{\cal A}^*$, (b) $r_{\cal B}^*$ as functions of $\phi$ for several values of $H$. $N=1800$, $L=100\sigma$, $H = -1\K$ and $K=0.1\K$.} 
  \label{fig:rinfty}
\end{figure}

%We calculated $t_c$ which is the average time it takes the oscillators to reach their almost complete synchronization, $r>0.95$.
%By plotting $t_c$ as a function of $\phi$, like in figure \ref{fig:t0KappaAlto}(a) it is possible to observe that increasing $\zeta$ decreases the value of $t_c$. If we plot $t_c$ as a function of $\zeta$, as in figure \ref{fig:t0KappaAlto}(b) we see that value of $t_c$ seems constant when $\zeta <1$ as the synchronization is mostly driven by the value of the highest coupling constant, that in this case is $K$. 

%\begin{figure}[h!]
  %  \centering
   % \includegraphics[width=1\linewidth]{t0KappaAlto.pdf}
 %   \caption{\textbf{Dependence of $t_c$ with $\phi$ and $\zeta$.} Panel (a) shows the dependence of $t_c$ on $\phi$ for various values of $\zeta$, by varying the value of $H$.  Panel (b) shows the dependence of $t_c$ on $\zeta$ for various values of $\phi$. For all cases $K=0.1$ and $N = 1800$.}
 %   \label{fig:t0KappaAlto}
%\end{figure}

\subsection{Model $\cal J$}

While model $\cal I$ is symmetric with respect to swapping the type of the oscillators and replacing $\phi$ by $1-\phi$, in model $\cal J$ this symmetry is lifted. ${\cal B}-$oscillators are contrarians to all $N$ particles in the system. Similar models, but on static random networks, were considered by several groups \cite{zanette2005,hong2011a,hong2011b,louzada2012,mirchev2014,ratas2016} with a common conclusion that it is possible to fully suppress the global synchronization, provided the number of contrarians or their negative coupling strength exceeds certain thresholds. Our results are in agreement with this general conclusion. Figure~\ref{fig:modelIIsynchronization} shows the time evolution of the order parameters $r_{\cal A}(t)$ and $r_{\cal B}(t)$ for both subpopulations at $\phi=0.5$ and for several values of the repulsive coupling constant $H$. Surprisingly, for low values of $|H|$ we observe an emergence of a partially synchronized states for contrarians with the late time values of the order parameter $r_{\cal B}>0.8$, see $H=-0.001$ curve in figure~\ref{fig:modelIIsynchronization}(b). Increasing the repulsive strength $|H|$ reduces the asymptotic value 
$r_{\sigma}^*\equiv r_{\sigma}(t \rightarrow \infty)$ for both subpopulations.

Next we discuss the dependence of  $r_{A}^*$ and $r_{B}^*$ upon  $H$, $K$ and $\phi$.  On Fig.~\ref{fig:diagramaA} we show heat maps of the order parameters $r_{A}^*$ and $r_{B}^*$ in the $(\phi,H)$ plane. The maps reveal the presence of continuous configurational transitions governed by either $H$ or $\phi$, at which $r_{\sigma}^*$, $\sigma = A,B$ decreases from 1 to 0 with decreasing $H$ at a certain fixed $\phi$, or with increasing $\phi$ at a certain fixed $H$.

In Fig.~\ref{fig:rinfty} we plot $r_{\cal A}^*$ and $r_{\cal B}^*$ for several cuts through the heat map at several fixed values of $H$. Curiously, $r_{\cal A}^*(\phi)$ [$r_{\cal B}^*(\phi)$] resembles the behavior of the spontaneous [field-driven] magnetization as a function of temperature in the Ising model, with $\phi$ playing the role of temperature. 
\vspace{-1cm}

\section{Conclusions}
\vspace{-0.5cm}

We have carried out extensive numerical simulations of the synchronization dynamics of two models of locally coupled mobile oscillators of two types. The first model deals with symmetric binary mixtures, where alike oscillators tend to synchronize, while unlike ones tend to be out of phase. The second model is asymmetric in this respect as it contains a given fraction of contrarians which tend do be out of phase with all the other oscillators in the system. 

We have focused in model $\cal I$ on the characteristic synchronization time $\tau$ describing the asymptotic exponential approach of the order parameter in Eq.~\eqref{eq:orderparameter} to unity, see also Figs.~\ref{fig:fitpanel}(b) and (d). We have found that $\tau$ decreases with increasing the composition $\phi$ of the mixture and for $|H| > K$, i.e., when the repulsive interactions are more intense as the attractive ones [Fig.~\ref{fig:KappaT0phi}(a) and Fig.~\ref{fig:tauf}(a)]. $\tau$ is also very sensitive to the ratio $\zeta = |H|/K$, and exhibits a power low decay for large enough $\zeta$ [Fig.~\ref{fig:tauf}(b)]. 
The synchronization dynamics proceed approximately as predicted by the mean-field theory, Eq.~(\ref{eq:kuramoto_r(t)}), in the fast switching regime which in our case is realized for $K$, and $|H| < 0.05 \K$ [Fig.~\ref{fig:fitpanel}(a), and Fig.~\ref{fig:binarysynchronizationpanel}(a)]. For larger values of the coupling constants, the dynamics of the phase synchronization can be mapped onto the coarsening relaxation dynamics of the $2D$ $XY$ model [Figs.~\ref{fig:binarysynchronizationpanel}(e)-(h) and Figs.~\ref{fig:states}], which in the intermediate times is dominated by the motion and annihilation of topological defect with opposite winding numbers. Controlling the overall number density of the oscillators provides additional means to enhance the synchronization, highlighting the role of the mobility of the oscillators [Fig.~\ref{fig:escalar}].
The general conclusion for model  $\cal I$ is the following: if a pure subpopulation exhibits a coherent attractor it will also be present in the mixtures.  

 In contrast, model ${\cal J}$ allows for a complete suppression of the coherent state, which can be achieved by increasing either $\zeta$ or $\phi$, as shown in the configuration diagrams in Fig.~\ref{fig:diagramaA}. For $K>0$, the late time values of both order parameters, $r_{\cal A}^*$, and $r_{\cal B}^*$ vary continuously between $1$ and $0$ with the increasing  of the mixture composition $\phi$ [Fig.~\ref{fig:rinfty}].  
 
 Here we have used a simplifying assumption that all the intrinsic frequencies $\omega_k^{\sigma}=0$, see Eq.~(\ref{eq:modkuramoto}). It is expected that the results obtained here will also hold for the case of non-zero, but uniform  frequencies $\omega_k^{\sigma}=\omega$. This however will not be the case if the frequencies are chosen at random from some distributions with densities $g^{\sigma}(\omega)$, $\sigma = {\cal A, B}$. Another possible extension of this work is to study the effects of coupling between the phases and the coordinates of the oscillators, e.g., by requiring that the oscillators move in the directions determined by their phase variables. Then the goal is to discover some novel swarming-like behaviors in the binary mixtures of self-propelled oscillators with the heterogeneous aligning interactions. One can also search for the conditions when binary systems 
 demix into ${\cal A}-$rich and ${\cal B}-$rich subpopulations, when for example the means of their respective $g^{\sigma}(\omega)$ have opposite signs.
\vspace{-0.5cm}

\section*{Acknowledgments}
\vspace{-0.5cm}

We acknowledge financial support from the Portuguese Foundation for Science and Technology (FCT) under Contracts no. PTDC/FIS-MAC/28146/2017 (LISBOA-01-0145-FEDER-028146), UIDB/00618/2020, UIDP/00618/2020, and IF/00322/2015.

\bibliography{sample}

\end{document}